\documentclass[11pt]{article}
\usepackage{a4wide}
\usepackage{amsmath,amssymb}
\usepackage{simplewick}
\usepackage{graphicx}
\usepackage{color}
\usepackage{braket}
\usepackage{comment}
\usepackage[bookmarks=true,bookmarksnumbered=true,setpagesize=false]{hyperref}

\DeclareMathOperator{\sgn}{sgn}
\DeclareMathOperator{\erf}{erf}
\DeclareMathOperator{\T}{T}

\usepackage{enumerate}
\newcommand{\al}[1]{\begin{align}#1\end{align}}
\newcommand{\als}[1]{\begin{align*}#1\end{align*}}
\newcommand{\ov}{\over}
\newcommand{\nn}{\nonumber\\}
\newcommand{\tx}{\text}

\newcommand{\paren}[1]{\left(#1\right)}
\newcommand{\sqbr}[1]{\left[#1\right]}
\newcommand{\ab}[1]{\left|#1\right|}

\newcommand{\fn}[1]{\!\left(#1\right)}

\newcommand{\Paren}[1]{\bigl(#1\bigr)}

\newcommand{\Fn}[1]{\!\bigl(#1\bigr)}

\newcommand{\bs}{\boldsymbol}

\newcommand{\df}{\text{d}}

\newcommand{\mc}{\mathcal}
\newcommand{\mf}{\mathfrak}
\newcommand{\bmf}[1]{\boldsymbol{\mathfrak#1}}

\newcommand{\p}{\partial}

\newcommand{\Or}[1]{\mathcal O\!\left(#1\right)}
\newcommand{\h}{\hat}
\newcommand{\ol}{\overline}
\newcommand{\pr}{\prime}

\newcommand{\commutator}[2]{\left[#1,#2\right]}

\newcommand{\wt}{\widetilde}

\newcommand{\vp}{\varphi}

\newcommand{\stin}{\varsigma_\tx{in}}
\newcommand{\stout}{\varsigma_\tx{out}}

\begin{document}
\title{Scalar scattering amplitude in Gaussian wave-packet formalism\bigskip\\}
\author{
Kenzo~Ishikawa,\footnote{
E-mail: \tt ishikawa@particle.sci.hokudai.ac.jp
}
\
Kenji~Nishiwaki,\footnote{
E-mail: \tt kenji.nishiwaki@snu.edu.in
}
\ 
and
Kin-ya~Oda\footnote{
E-mail: \tt odakin@phys.sci.osaka-u.ac.jp
}\bigskip\\
$^*$ \it Department of Physics, 
Hokkaido University, Hokkaido 060-0810, Japan \\
$^*$\it\normalsize Research and Education Center for Natural Sciences, Keio University Kanagawa 223-8521, Japan\\
$^\dagger$ \it\normalsize Department of Physics, 
Shiv Nadar University, Gautam Buddha
Nagar 201314, India \\
$^\ddag$ \it\normalsize Department of Physics, 
Osaka University, Osaka 560-0043, Japan
}
\maketitle


\begin{abstract}\noindent
We compute an $s$-channel $2\to2$ scalar scattering $\phi\phi\to\Phi\to\phi\phi$ in the Gaussian wave-packet formalism at the tree-level.
We find that wave-packet effects, including shifts of the pole and width of the  propagator of $\Phi$, persist even when we do not take into account the time-boundary effect for $2\to2$, proposed earlier.
The result can be interpreted that a heavy scalar $1\to2$ decay $\Phi\to\phi\phi$, taking into account the production of $\Phi$, does not exhibit the in-state time-boundary effect unless we further take into account in-boundary effects for the $2\to2$ scattering. We also show various plane-wave limits.
\end{abstract}
\newpage

\tableofcontents

\newpage

\section{Introduction and summary}
It is well-known that a plane-wave S-matrix is ill-defined when taken literally because its matrix element is proportional to the energy-momentum delta function, which always gives either zero or infinity when squared to compute a probability.
On the other hand, we may define an S-matrix in the Gaussian wave-packet basis without such an infinity~\cite{Ishikawa:2005zc,Ishikawa:2018koj}.

It has been claimed that the Gaussian formalism gives a deviation from the Fermi's golden rule~\cite{Ishikawa:2013kba,Ishikawa:2016lnn}, in which the probability is suppressed only by a power of the deviation from the energy-momentum conservation rather than the conventional exponential suppression;\footnote{
One might find relevance to the use of the crystal ball function; see e.g.\ Appendix F in Ref.~\cite{Gaiser:1982yw}.
}
see also Refs.~\cite{Ishikawa:2014eoa,Ishikawa:2014uma,Fujii:2014vaa}.

In Ref.~\cite{Ishikawa:2018koj}, a scalar decay $\Phi\to\phi\phi$ has been computed in the Gaussian formalism, and the previously-claimed power-law deviation from the Fermi's golden rule has been identified to come from the configuration in which the decay interaction is placed near a time-boundary.
As we will see, this configuration is realized, even if the in/out states are at a distance.
To examine the in-boundary effect for $1\to2$ more in detail, it is desirable to take into account the production process of the decaying $\Phi$.

In this paper, we compute a tree-level $s$-channel scalar scattering $\phi\phi\to\Phi\to\phi\phi$ in the Gaussian formalism.
We find that wave-packet effects, including shifts of the pole and width of the  propagator of $\Phi$, persists even when we do not take into account the time-boundary effect, proposed earlier.
The result can be interpreted that a heavy scalar decay $\Phi\to\phi\phi$, taking into account the production of $\Phi$, does not exhibit the in-state $1\to2$ time-boundary effect unless we take into account the in-state $2\to2$ time boundary.

This paper is organized as follows: In Sec.~\ref{Gaussian section}, we present basic setup of the Gaussian formalism, and compute the Gaussian S-matrix for the $s$-channel $2\to2$ scattering: $\phi\phi\to\Phi\to\phi\phi$.
In Sec.~\ref{boundary effect section}, we discuss the possible time-boundary effects. In Sec.~\ref{Bulk amplitude section}, we focus on the bulk contribution and show that wave effects exist even when we neglect the boundary contributions. In Sec.~\ref{limits section}, we present several plane-wave limits of the obtained result. In Sec.~\ref{summary section}, we present summary and discussion. In Appendix~\ref{phi 4 section}, we compare with the $\phi\phi\to\phi\phi$ scattering in the $\phi^4$ theory.

\section{Gaussian S-matrix}\label{Gaussian section}
Here we first review the Gaussian formalism, and obtain the S-matrix for the $s$-channel $2\to2$ scalar scattering: $\phi\phi\to\Phi\to\phi\phi$.

\subsection{Gaussian basis}
We review the Gaussian formalism, following Ref.~\cite{Ishikawa:2018koj}, to clarify the notation in this paper.
A free scalar field operator $\h\vp$ at $x=\paren{x^0,\bs x}$ (in the interaction picture) can be expanded by the plane basis:
\al{
\h\vp\fn{x}
	&=	\left.\int{\df^3\bs p\ov\sqrt{2p^0}\paren{2\pi}^{3/2}}\sqbr{e^{ip\cdot x}\h a_\vp\fn{\bs p}+\tx{h.c.}}\right|_{p^0=E_\vp\fn{\bs p}}\nn
	&=	\left.\int{\df^3\bs p\ov\sqrt{2p^0}}\sqbr{\Braket{\vp;x|\vp;\bs p}\h a_\vp\fn{\bs p}+\tx{h.c.}}\right|_{p^0=E_\vp\fn{\bs p}},
		\label{free fields}
}
where $\vp=\phi,\Phi$ labels the particle species;
$\h a_\vp\fn{\bs p}$ and $\h a_\vp^\dagger\fn{\bs p}$ are the annihilation and creation operators, respectively, with
\al{
\commutator{\h a_\vp\fn{\bs p}}{\h a_{\vp'}^\dagger\fn{\bs p'}}
	&=	\delta_{\vp\vp'}\delta^3\fn{\bs p-\bs p'},&
\tx{others}
	&=	0;
}
and
\al{
E_\vp\fn{\bs p}
	&:=	\sqrt{m_\vp^2+\bs p^2},\\
\Ket{\vp;\bs p}
	&:=	\h a^\dagger_\vp\fn{\bs p}\Ket{0},\\
\Braket{\vp;\bs x|\vp';\bs p}
	&:=	\delta_{\vp\vp'}{e^{i\bs p\cdot\bs x}\ov\paren{2\pi}^{3/2}},\\
\Ket{\vp;x}
	&:=	e^{+i\h H_\tx{free}t}\Ket{\vp;\bs x},
}
with $\h H_\tx{free}$ being the free Hamiltonian:
\al{
\h H_\tx{free}\Ket{\vp;\bs p}
	&=	E_\vp\fn{\bs p}\Ket{\vp;\bs p}.
}
Here and hereafter, we use $t,T$ and $x^0,X^0$ interchangeably: $t=x^0$ and $T=X^0$.
Note that $\Ket{\varphi;\bs x}$ and $\Ket{\varphi;\bs p}$ are independent of time and hence can be regarded as either a Heisenberg-picture  state or a Schr\"odinger-picture eigenbasis (of total Hamiltonian), while $\Ket{\varphi;x}$ is an interaction-picture basis at time $x^0$ as seen from its time evolution by the free Hamiltonian.

We define a Gaussian wave-packet state $\Ket{\varphi,\sigma;\bs\Pi}$ by
\al{
\Braket{\varphi',\bs x|\varphi,\sigma;\bs\Pi}
	&:=	{1\ov\paren{\pi\sigma}^{3/4}}e^{i\bs P\cdot\paren{\bs x-\bs X}}e^{-{1\ov2\sigma}\paren{\bs x-\bs X}^2}\delta_{\varphi\varphi'},
	\label{Gaussian wave function}
}
where $\bs\Pi:=\paren{\bs X,\bs P}$ gives the center of wave packet in the phase space. Note that
\al{
\Braket{\varphi',\bs p|\varphi,\sigma;\bs\Pi}
	&=	\delta_{\varphi\varphi'}\paren{\sigma\ov\pi}^{3/4}e^{-i\bs p\cdot\bs X}e^{-{\sigma\ov2}\paren{\bs p-\bs P}^2},\\
\Braket{\varphi,\sigma;\bs\Pi|\varphi',\sigma';\bs\Pi'}
	&=	\paren{\sigma_\tx{I}\ov\sigma_\tx{A}}^{3/4}
		e^{-{1\ov4\sigma_\tx{A}}\paren{\bs X-\bs X'}^2}
		e^{-{\sigma_\tx{I}\ov4}\paren{\bs P-\bs P'}^2}
		e^{{i\ov2\sigma_\tx{I}}\paren{\sigma\bs P+\sigma'\bs P'}\cdot\paren{\bs X-\bs X'}}\delta_{\varphi\varphi'},
}
where
\al{
\sigma_\tx{A}
	&:=	{\sigma+\sigma'\ov2},&
\sigma_\tx{I}
	&:=	\paren{\sigma^{-1}+\sigma^{\pr-1}\ov2}^{-1}
	=	{2\sigma\sigma'\ov\sigma+\sigma'},
}
are the average and the inverse of average of inverse, respectively.
Especially,
\al{
\Braket{\varphi,\sigma;\bs\Pi|\varphi,\sigma;\bs\Pi'}
	&=	e^{-{1\ov4\sigma}\paren{\bs X-\bs X'}^2}
		e^{-{\sigma\ov4}\paren{\bs P-\bs P'}^2}
		e^{{i\ov2}\paren{\bs P+\bs P'}\cdot\paren{\bs X-\bs X'}}.
}

The state $\Ket{\varphi,\sigma;\bs\Pi}$ is time independent and hence can be regarded as either a Heisenberg state or a Schr\"odinger basis. 
We also define the interaction basis at time $X^0$:
\al{
\Ket{\varphi,\sigma;\Pi}
	&:=	e^{i\h H_\tx{free}X^0}\Ket{\varphi,\sigma;\bs\Pi},
		\label{relation between 3d and 4d}
}
where $\Pi:=\paren{X,\bs P}=\paren{X^0,\bs X,\bs P}=\paren{X^0,\bs\Pi}$.
As we will see later, we will treat $\Ket{\varphi,\sigma;\Pi}$ as a time-independent Heisenberg state (or equivalently a time-independent Schr\"odinger basis).

We define a creation operator of the Gaussian basis by
\al{
\h A_{\varphi,\sigma}^\dagger\fn{\Pi}\Ket{0}
	&:=	\Ket{\varphi,\sigma;\Pi},
}
which results in $\h A_{\varphi,\sigma}\fn{\Pi}\Ket{0}=0$ and
\al{
\commutator{\h A_{\varphi,\sigma}\fn{\Pi}}{\h A^\dagger_{\varphi',\sigma'}\fn{\Pi'}}
	&=	\Braket{\varphi,\sigma,\Pi|\varphi',\sigma';\Pi'},&
\tx{others}
	&=	0.
}
We may also expand $\h\vp$ by the creation and annihilation operators of the free Gaussian wave packets:
\al{
\h\vp\fn{x}
	&=	\int{\df^3\bs X\,\df^3\bs P\ov\paren{2\pi}^3}
			\sqbr{f_{\vp,\sigma;X,\bs P}\fn{x}\h A_{\vp,\sigma}\fn{X,\bs P}
					+\tx{h.c.}},
}
where 
$X=\paren{X^0,\bs X}$ is the center of the wave packet;
$\bs P$ is the central momentum of the wave packet; 
 $\sigma$ and $X^0$ are fixed (and can differ) for each field participating in the scattering;
and 
the coefficient function becomes
\al{
f_{\vp,\sigma;X,\bs P}\fn{x}
	&:=	\int{\df^3\bs p\ov\sqrt{2E_\vp\fn{\bs p}}}\Braket{\vp;x|\vp;\bs p}\Braket{\vp;\bs p|\vp,\sigma;\Pi}\nn
	&=	\left.\paren{\sigma\ov\pi}^{3/4}\int{\df^3\bs p\ov\sqrt{2p^0}\paren{2\pi}^{3/2}}e^{ip\cdot\paren{x-X}-{\sigma\ov2}\paren{\bs p-\bs P}^2}\right|_{p^0=E_\vp\fn{\bs p}}.
}
We also write
\al{
\df^6\bs\Pi
	&:=	{\df^3\bs X\,\df^3\bs P\ov\paren{2\pi}^3}
}
so that
\al{
\h\vp\fn{x}
	&=	\int\df^6\bs\Pi
			\sqbr{f_{\vp,\sigma;\Pi}\fn{x}\h A_{\vp,\sigma}\fn{\Pi}
					+\tx{h.c.}}.
		\label{Gaussian expansion}
}
By e.g.\ sandwiching between $\Bra{\bs p}$ and $\Ket{\bs p'}$, we can show the completeness of the Gaussian basis in the one-particle subspace:
\al{
\int\df^6\bs\Pi\Ket{\varphi,\sigma;\Pi}\Bra{\varphi,\sigma;\Pi}
	&=	\h1.
}
Namely, the Gaussian basis can expand any one-particle wave function $\psi\fn{\bs x}=\Braket{\bs x|\psi}$ as
\al{
\Braket{\bs x|\psi}
	&=	\int\df^6\bs\Pi\Braket{\bs x|\bs\Pi}\Braket{\bs\Pi|\psi},
}
where we used the short-hand notation $\Ket{\bs\Pi}=\Ket{\varphi,\sigma;\bs\Pi}$ etc.\ and $\Braket{\bs x|\bs\Pi}$ is given in Eq.~\eqref{Gaussian wave function}.
We have also used $\Ket{\Pi}\Bra{\Pi}=\Ket{\bs\Pi}\Bra{\bs\Pi}$ from Eq.~\eqref{relation between 3d and 4d}.
Note the following relation:
\al{
\Bra{0}\h A_{\vp,\sigma}\fn{\Pi}\h A_{\vp',\sigma'}^\dagger\fn{\Pi'}\Ket{0}
	&=	\Braket{\vp,\sigma;\Pi|\vp,\sigma';\Pi'}\delta_{\vp\vp'},\\
\left.\Braket{\vp,\sigma;\Pi|\vp,\sigma;\Pi'}\right|_{X^0=X^{\pr0}}
	&=	e^{-{1\ov4\sigma}\paren{\bs X-\bs X'}^2}
		e^{-{\sigma\ov4}\paren{\bs P-\bs P'}^2}
		e^{{i\ov2}\paren{\bs P+\bs P'}\cdot\paren{\bs X-\bs X'}}.
		\label{Gaussian inner product}
}

In the large-$\sigma$ expansion, we get
\al{
f_{\vp,\sigma;X,\bs P}\fn{x}
	&\to
		\left.\paren{\sigma\ov\pi}^{3/4}\paren{2\pi\ov\sigma}^{3/2}
			{1\ov\sqrt{2P^0}\paren{2\pi}^{3/2}}e^{iP\cdot\paren{x-X}-{\paren{\bs x-\bs\Xi_\vp^\Pi\fn{x^0}}^2\ov2\sigma}}\right|_{P^0=E_\vp\fn{\bs P}},
			\label{plane-wave limit of f}
}
where
\al{
\bs\Xi^\Pi_{\vp}\fn{x^0}
	&:=	\bmf X_\varphi^\Pi+\bs V_\vp\fn{\bs P}x^0\nn
	&=	\bs X+\bs V_\vp\fn{\bs P}\paren{x^0-X^0},
}
in which
\al{
\bmf X_\varphi^\Pi
	&:=	\bs X-\bs V_\vp\fn{\bs P}X^0,&
\bs V_\vp\fn{\bs P}
	&:=	{\bs P\ov E_\vp\fn{\bs P}}.
}

\subsection{In and out states}
We consider the $s$-channel scalar scattering $\phi\phi\to\Phi\to\phi\phi$. Since both the in and out states are of $\phi$, we omit the label $\phi$ hereafter.

Generically, one particle in the in- and out-states can be asymptotic to an arbitrary free wave function $\Psi\fn{x}=\Braket{x|\Psi}$, which can be expanded by the Gaussian basis as
\al{
\Ket{\Psi}
	&=	\int\df^6\bs\Pi\,\Ket{\bs\Pi}\Braket{\bs\Pi|\Psi}.
}
Therefore without loss of generality, we may assume that the asymptotic free states are Gaussian, and we will do so hereafter.

We prepare the in and out Heisenberg states $\Ket{\tx{in};\sigma_1,\Pi_1;\sigma_2,\Pi_2}$ and  $\Ket{\tx{out};\sigma_3,\Pi_3;\sigma_4,\Pi_4}$, respectively, by
\al{
e^{-i\h Ht}\Ket{\tx{in};\sigma_1,\Pi_1;\sigma_2,\Pi_2}
	&\to	e^{-i\h H_\tx{free}t}\Ket{\sigma_1,\Pi_1;\sigma_2,\Pi_2}&
(t&\to T_\tx{in}),\nn
e^{-i\h Ht}\Ket{\tx{out};\sigma_3,\Pi_3;\sigma_4,\Pi_4}
	&\to	e^{-i\h H_\tx{free}t}\Ket{\sigma_3,\Pi_3;\sigma_4,\Pi_4}&
(t&\to T_\tx{out}),
	\label{boundary relation}
}
where we have defined the free states
\al{
\Ket{\sigma_1,\Pi_1;\sigma_2,\Pi_2}
	&:=	\h A^\dagger_{\sigma_1}\fn{\Pi_1}\h A^\dagger_{\sigma_2}\fn{\Pi_2}\Ket{0},
}
etc., and take
\al{
T_\tx{in}
	&\lesssim
		\max\fn{X_1^0,X_2^0},&
T_\tx{out}
	&\gtrsim
		\min\fn{X_3^0,X_4^0}.
}
See Sec.~\ref{boundary effect section} for further discussion.

\subsection{Gaussian two-point function}
In this subsection, we omit the labels $\vp$ and $\sigma$ as they are all equal, except for the mass $m_\varphi$. In the later application, $\varphi$ will be the intermediate heavy scalar $\Phi$.

We want to put the expansion~\eqref{Gaussian expansion},
into the time-ordered two-point function:
\al{
\Bra{0}\T\h\vp\fn{x}\h\vp\fn{x'}\Ket{0}
	&=	\theta\fn{x^0-x^{\pr0}}
		\Bra{0}\h\vp\fn{x}\h\vp\fn{x'}\Ket{0}
		+\theta\fn{x^{\pr0}-x^0}
		\Bra{0}\h\vp\fn{x'}\h\vp\fn{x}\Ket{0}.
		\label{two-point function}
}
Now we can check that
\al{
\Bra{0}\h\vp\fn{x}\h\vp\fn{x'}\Ket{0}
	&=	\lefteqn{\int\df^6\bs\Pi\,\int\df^6\bs\Pi'\,
		f_{\Pi}\fn{x}
		f^*_{\Pi'}\fn{x'}
		\Bra{0}\h A\fn{\Pi}\h A^\dagger\fn{\Pi'}\Ket{0}}\nn
	&=	\int{\df^3\bs p\ov\sqrt{2E\fn{\bs p}}}\int{\df^3\bs p'\ov\sqrt{2E\fn{\bs p'}}}\int\df^6\bs\Pi\,\int\df^6\bs\Pi'\nn
	&\quad\times
		\Braket{x|\bs p}\Braket{\bs p|\Pi}\Braket{\Pi|\Pi'}\Braket{\Pi'|\bs p'}
		\Braket{\bs p'|x'}\nn
	&=	\int{\df^3\bs p\ov2E\fn{\bs p}}
		\Braket{x|\bs p}
		\Braket{\bs p|x'}
	=	\left.\int{\df^3\bs p\ov2E\fn{\bs p}\paren{2\pi}^3}
		e^{ip\cdot\paren{x-x'}}\right|_{p^0=E\fn{\bs p}}.
}
Putting this into the two-point function~\eqref{two-point function},
\al{
\Bra{0}\T\h\vp\fn{x}\h\vp\fn{x'}\Ket{0}
	&=	\left.
		\int{\df^3\bs p\ov2E\fn{\bs p}\paren{2\pi}^3}
		\paren{
			\theta\fn{x^0-x^{\pr0}}
				e^{ip\cdot\paren{x-x'}}
			+\theta\fn{x^{\pr0}-x^0}
				e^{ip\cdot\paren{x'-x}}
			}
		\right|_{p^0=E\fn{\bs p}}.
}
We have recovered the ordinary plane-wave propagator as we should, since we integrate over the complete set.\footnote{
See Ref.~\cite{Feynman:1949hz} for an early work by Feynman containing consideration with waves.
}
As usual, using
\al{
\theta\fn{x^0}
	&=	-{1\ov2\pi i}\int_{-\infty}^\infty\df\omega {e^{-i\omega x^0}\ov\omega+i\epsilon},
}
with $\epsilon$ being an arbitrary positive infinitesimal, we may rewrite it into more familiar form:
\al{
\Bra{0}\T\h\vp\fn{x}\h\vp\fn{x'}\Ket{0}
	&=	\int{\df^3\bs p\,e^{i\bs p\cdot\paren{\bs x-\bs x'}}\ov2E\fn{\bs p}\paren{2\pi}^3}\nn
	&\quad\times
		\paren{
			-\int_{-\infty}^\infty{\df\omega\ov2\pi i} {e^{-i\paren{\omega+E\fn{\bs p}}\paren{x^0-x^{\pr0}}}\ov\omega+i\epsilon}
			-\int_{-\infty}^\infty{\df\omega\ov2\pi i} {e^{-i\paren{\omega-E\fn{\bs p}}\paren{x^0-x^{\pr0}}}\ov-\omega+i\epsilon}
			}\nn
	&=	{i\ov\paren{2\pi}^4}\int{\df^3\bs p\ov 2E\fn{\bs p}}\int_{-\infty}^\infty\df p^0\,
		e^{ip\cdot\paren{x-x'}}\nn
	&\quad\times
		\paren{
			{1\ov p^0-E\fn{\bs p}+i\epsilon}
			+{1\ov-p^0-E\fn{\bs p}+i\epsilon}
			}\nn
	&=	{i\ov\paren{2\pi}^4}\int\df^3\bs p\int_{-\infty}^\infty\df p^0\,
		e^{ip\cdot\paren{x-x'}}
		{
			-1
			\ov\paren{\bs p^2+m_\vp^2-i\epsilon}-\paren{p^0}^2
			}\nn
	&=	-i\int{\df^4p\ov\paren{2\pi}^4}{e^{ip\cdot\paren{x-x'}}\ov p^2+m_\varphi^2-i\epsilon}
	=	-i\Delta_\tx{F}\fn{x-x'}.
}

\subsection{Gaussian S-matrix}

Now we compute the probability amplitude under the assumption~\eqref{boundary relation}:
\al{
\mc S
	&=	\Braket{\tx{out};\sigma_3,\Pi_3;\sigma_4;\Pi_4|\tx{in};\sigma_1,\Pi_1;\sigma_2,\Pi_2}\nn
	&=	\Bra{\sigma_3,\Pi_3;\sigma_4,\Pi_4}e^{i\h H_\tx{free}T_\tx{out}}e^{-i\h HT_\tx{out}}e^{i\h HT_\tx{in}}e^{-i\h H_\tx{free}T_\tx{in}}\Ket{\sigma_1,\Pi_1;\sigma_2,\Pi_2}\nn
	&=	\Bra{\sigma_3,\Pi_3;\sigma_4,\Pi_4}
		\T\exp\fn{-i\int_{T_\tx{in}}^{T_\tx{out}}\df t\,\h H^\tx{I}_\tx{int}\fn{t}}
		\Ket{\sigma_1,\Pi_1;\sigma_2,\Pi_2}\nn
	&=:	\Bra{\sigma_3,\Pi_3;\sigma_4,\Pi_4}
		\h S
		\Ket{\sigma_1,\Pi_1;\sigma_2,\Pi_2},
			\label{Dyson series}
}
where $\h H^\tx{I}_\tx{int}\fn{t}=e^{i\h H_\tx{free}t}\paren{\h H-\h H_\tx{free}}e^{-i\h H_\tx{free}t}$ is the interaction Hamiltonian in the interaction picture.
In the plane-wave S-matrix, one subtracts the first term in the Dyson series~\eqref{Dyson series}, write $\h S=\h1+i\h T$, and concentrate on the transition amplitude from $\h T$. In the Gaussian formalism, we do not need such regularization of dropping the first term $\h1$ because the inner product of the free states would remain finite even for identical momenta.\footnote{
Recall Eq.~\eqref{Gaussian inner product} for an explicit formula for particular equal-time packets.
}
When we integrate over the final~state momenta $\bs P_3$ and $\bs P_4$, the contribution from $\h1$ would automatically drop out even if we take the plane-wave limit after all the computations. Hereafter, we omit the trivial term $\Braket{\sigma_3,\Pi_3;\sigma_4,\Pi_4|\sigma_1,\Pi_1;\sigma_2,\Pi_2}$ from $\mc S$ when we call it ``transition amplitude''.

In this paper, we consider the following simplest interaction Hamiltonian:
\al{
\h H_\tx{int}^\tx{I}\fn{t}
	&=	{\kappa\ov2}\int\df^3\bs x\,\h\phi^2\fn{x}\h\Phi\fn{x},
}
where $\h\phi$ and $\h\Phi$ are given in Eq.~\eqref{free fields}.
The tree-level transition amplitude is given by
\al{
\mc S
	&=	{\paren{-i\kappa}^2\ov8}
		\int_{T_\tx{in}}^{T_\tx{out}}\df t\int\df^3\bs x
		\int_{T_\tx{in}}^{T_\tx{out}}\df t'\int\df^3\bs x'\nn
	&\quad\times
		\Bra{0}\T_{x,x'}
		\h A_{\sigma_3}\fn{\Pi_3}\h A_{\sigma_4}\fn{\Pi_4}
		\h\phi\fn{x}\h\phi\fn{x}\h\Phi\fn{x}
		\h\phi\fn{x'}\h\phi\fn{x'}\h\Phi\fn{x'}
		\h A^\dagger_{\sigma_1}\fn{\Pi_1}\h A^\dagger_{\sigma_2}\fn{\Pi_2}
		\Ket{0},
}
where $\T_{x,x'}$ is the time ordering with respect to $x$ and $x'$ only.
Hereafter, we concentrate on the $s$-channel process because it is dominant in the near on-shell process of our interest.

For example, a part of the $s$-channel process is
\al{
\mc S	
	&\supset
		{\paren{-i\kappa}^2\ov8}
		\int_{T_\tx{in}}^{T_\tx{out}}\df t\int\df^3\bs x
		\int_{T_\tx{in}}^{T_\tx{out}}\df t'\int\df^3\bs x'\nn
	&\quad\times\Bra{0}T_{x,x'}
\contraction{}{\h A}
	{_{\sigma_3}\fn{\Pi_3}\h A_{\sigma_4}\fn{\Pi_4}}{\h\phi}
\contraction[2ex]{\h A_{\sigma_3}\fn{\Pi_3}}{\h A}
	{_{\sigma_4}\fn{\Pi_4}\h\phi\fn{x}}{\h\phi}
\bcontraction
	{\h A_{\sigma_3}\fn{\Pi_3}\h A_{\sigma_4}\fn{\Pi_4}\h\phi\fn{x}\h\phi\fn{x}}
	{\h\Phi}{\fn{x}\h\phi\fn{x'}\h\phi\fn{x'}}{\h\Phi}
\contraction
	{\h A_{\sigma_3}\fn{\Pi_3}\h A_{\sigma_4}\fn{\Pi_4}\h\phi\fn{x}\h\phi\fn{x}\h\Phi\fn{x}}
	{\h\phi}{\fn{x'}\h\phi\fn{x'}\h\Phi\fn{x'}}
	{\h A}
\contraction[2ex]
	{\h A_{\sigma_3}\fn{\Pi_3}\h A_{\sigma_4}\fn{\Pi_4}
		\h\phi\fn{x}\h\phi\fn{x}\h\Phi\fn{x}
		\h\phi\fn{x'}}
	{\h\phi}
	{\fn{x'}\h\Phi\fn{x'}
		\h A^\dagger_{\sigma_1}\fn{\Pi_1}}
	{\h A}
		\h A_{\sigma_3}\fn{\Pi_3}\h A_{\sigma_4}\fn{\Pi_4}
		\h\phi\fn{x}\h\phi\fn{x}\h\Phi\fn{x}
		\h\phi\fn{x'}\h\phi\fn{x'}\h\Phi\fn{x'}
		\h A^\dagger_{\sigma_1}\fn{\Pi_1}\h A^\dagger_{\sigma_2}\fn{\Pi_2}
		\Ket{0}.
		\label{a contribution}
}
The Wick contraction with the external line gives, for example,
\al{
\contraction{}{\h A}{_{\sigma_3}\fn{\Pi_3}}{\h\phi}
\h A_{\sigma_3}\fn{\Pi_3}\h\phi\fn{x}
	&=	\int\df^6\bs\Pi f^*_{\sigma;\Pi}\fn{x}
		\commutator{\h A_{\sigma_3}\fn{\Pi_3}}
			{\h A^\dagger_{\sigma}\fn{\Pi}}\nn
	&=	\int\df^6\bs\Pi
		\int{\df^3\bs p\ov\sqrt{2E_\phi\fn{\bs p}}}
			\Braket{\sigma;\Pi|\phi,\bs p}\Braket{\phi,\bs p|\phi,x}
			\Braket{\sigma_3;\Pi_3|\phi,\sigma;\Pi}\nn
	&=	\int{\df^3\bs p\ov\sqrt{2E_\phi\fn{\bs p}}}
			\Braket{\sigma_3;\Pi_3|\phi,\bs p}
			\Braket{\phi,\bs p|\phi,x}\nn
	&=	f^*_{\sigma_3;\Pi_3}\fn{x},
}
where the propagator of $\Phi$ becomes the same as the plane-wave one, as we have seen in the previous sub-section.
Then the contribution~\eqref{a contribution} becomes
\al{
\mc S
	&\supset
		{\paren{-i\kappa}^2\ov8}
		\int_{T_\tx{in}}^{T_\tx{out}}\df t\int\df^3\bs x
		\int_{T_\tx{in}}^{T_\tx{out}}\df t'\int\df^3\bs x'\nn
	&\quad\times
		f_{\sigma_1;\Pi_1}\fn{x'}
		f_{\sigma_2;\Pi_2}\fn{x'}
		f^*_{\sigma_3;\Pi_3}\fn{x}
		f^*_{\sigma_4;\Pi_4}\fn{x}
		\Bra{0}\T
		\bcontraction{}{\h\Phi}{\fn{x}}{\h\Phi}
		\h\Phi\fn{x}\h\Phi\fn{x'}\Ket{0}.
}
In total there will be factor 8 from the other Wick contractions.
To summarize,
\al{
\mc S
	&=	\paren{-i\kappa}^2
		\paren{-i}\int{\df^4p\ov\paren{2\pi}^4}{1\ov p^2+M^2-i\epsilon}\nn
	&\quad\times
		\int_{T_\tx{in}}^{T_\tx{out}}\df t\int\df^3\bs x\,
		f^*_{\sigma_3;\Pi_3}\fn{x}
		f^*_{\sigma_4;\Pi_4}\fn{x}
		e^{ip\cdot x}\nn
	&\quad\times
		\int_{T_\tx{in}}^{T_\tx{out}}\df t'\int\df^3\bs x'\,
		f_{\sigma_1;\Pi_1}\fn{x'}
		f_{\sigma_2;\Pi_2}\fn{x'}
		e^{-ip\cdot x'},
}
where $t:=x^0$ and $t':=x^{\pr0}$ are the production and decay times of $\Phi$, and $M:=m_\Phi$ is the heavy scalar mass.
This is the starting equation for our computation.

Hereafter, we consider the leading approximation in the plane-wave limit~\eqref{plane-wave limit of f}:
\al{
f_{\phi,\sigma_1;\Pi_1}\fn{x}
f_{\phi,\sigma_2;\Pi_2}\fn{x}
&\to	
		\paren{1\ov\pi\sigma_1}^{3/4}\paren{1\ov\pi\sigma_2}^{3/4}
		{1\ov\sqrt{2E_1}\sqrt{2E_2}}\nn
&\quad\times
		e^{iP_1\cdot\paren{x-X_1}-{\paren{\bs x-\bs\Xi_1\fn{t}}^2\ov2\sigma_1}}
		e^{iP_2\cdot\paren{x-X_2}-{\paren{\bs x-\bs\Xi_2\fn{t}}^2\ov2\sigma_2}},\nn
f^*_{\phi,\sigma_3;\Pi_3}\fn{x}
f^*_{\phi,\sigma_4;\Pi_4}\fn{x}
&\to	
		\paren{1\ov\pi\sigma_3}^{3/4}\paren{1\ov\pi\sigma_4}^{3/4}
		{1\ov\sqrt{2E_3}\sqrt{2E_4}}\nn
&\quad\times
		e^{-iP_3\cdot\paren{x-X_3}-{\paren{\bs x-\bs\Xi_3\fn{t}}^2\ov2\sigma_3}}
		e^{-iP_4\cdot\paren{x-X_4}-{\paren{\bs x-\bs\Xi_4\fn{t}}^2\ov2\sigma_4}},
}
where for $a=1,\dots,4$,
\al{
\bs\Xi_a\fn{t}
	&:=	\bmf X_a+\bs V_at,
}
in which $\bmf X_a$ is the center of wave packet at a reference time $t=0$ and $\bs V_a$ is its central velocity:
\al{
\bmf X_a
	&:=	\bs X_a-\bs V_aT_a,\\
\bs V_a
	&:=	{\bs P_a\ov E_a}
	=	{\bs P_a\ov\sqrt{m^2+\bs P_a^2}},
}
with $m:=m_\phi$. 

We perform the Gaussian integral over the positions of interaction to get
\al{
\mc S	
	&=	i\kappa^2\paren{
			\prod_{A=1}^4{1\ov\sqrt{2E_A}}
			\paren{1\ov\pi\sigma_A}^{3/4}}
		\paren{2\pi\sigma_\tx{in}}^{3/2}
		\paren{2\pi\sigma_\tx{out}}^{3/2}
		\int{\df^4p\ov\paren{2\pi}^4}{1\ov p^2+M^2-i\epsilon}
		\nn
	&\quad\times
		\int_{T_\tx{in}}^{T_\tx{out}} \df t\,e^{
			-{\sigma_\tx{out}\ov2}\paren{\bs p-\bs P_\tx{out}}^2
			-{1\ov2\stout}\paren{t-\mf T_\tx{out}}^2
			-{\mc R_\tx{out}\ov2}
			-it\paren{p^0-E_\tx{out}}
			+i\ol{\bs V}_\tx{out}\cdot\paren{\bs p-\bs P_\tx{out}}t
			+i\ol{\bmf X}_\tx{out}\cdot\paren{\bs p-\bs P_\tx{out}}}\nn		
	&\quad\times
		\int_{T_\tx{in}}^{T_\tx{out}} \df t'\,e^{
			-{\sigma_\tx{in}\ov2}\paren{\bs p-\bs P_\tx{in}}^2
			-{1\ov2\stin}\paren{t'-\mf T_\tx{in}}^2
			-{\mc R_\tx{in}\ov2}
			+it'\paren{p^0-E_\tx{in}}
			-i\ol{\bs V}_\tx{in}\cdot\paren{\bs p-\bs P_\tx{in}}t'
			-i\ol{\bmf X}_\tx{in}\cdot\paren{\bs p-\bs P_\tx{in}}},
			\label{S master}
}
where we have dropped a phase factor that cancels out in the square $\ab{S}^2$ and have defined the following:
\begin{itemize}
\item Energies and momanta for in and out states:
\al{
E_\tx{in}
	&:=	E_1+E_2,&
\bs P_\tx{in}
	&:=	\bs P_1+\bs P_2,\\
E_\tx{out}
	&:=	E_3+E_4,&
\bs P_\tx{out}
	&:=	\bs P_3+\bs P_4.
}
\item The averaged space-like width-squared of the in- and out-states, respectively:
\al{
\sigma_\tx{in}
	&:=	\paren{{1\ov\sigma_1}+{1\ov\sigma_2}}^{-1},&
\sigma_\tx{out}
	&:=	\paren{{1\ov\sigma_3}+{1\ov\sigma_4}}^{-1}.
}
\item For any three vector $\bs Q$,
\al{
\ol{\bs Q}_\tx{in}
	&:=	\sigma_\tx{in}\paren{{\bs Q_1\ov\sigma_1}+{\bs Q_2\ov\sigma_2}},&
\ol{\bs Q}_\tx{in}^2
	&:=	\ol{\bs Q}_\tx{in}\cdot\ol{\bs Q}_\tx{in},&
\ol{\bs Q^2}_\tx{in}
	&:=	\sigma_\tx{in}\paren{{\bs Q_1^2\ov\sigma_1}+{\bs Q_2^2\ov\sigma_2}},\\
\ol{\bs Q}_\tx{out}
	&:=	\sigma_\tx{out}\paren{{\bs Q_3\ov\sigma_3}+{\bs Q_4\ov\sigma_4}},&
\ol{\bs Q}_\tx{out}^2
	&:=	\ol{\bs Q}_\tx{out}\cdot\ol{\bs Q}_\tx{out},&
\ol{\bs Q^2}_\tx{out}
	&:=	\sigma_\tx{out}\paren{{\bs Q_3^2\ov\sigma_3}+{\bs Q_4^2\ov\sigma_4}},
}
and
\al{
\Delta\bs Q^2_\tx{in}
	&:=	\ol{\bs Q^2}_\tx{in}-\ol{\bs Q}_\tx{in}^2,&
\Delta\bs Q^2_\tx{out}
	&:=	\ol{\bs Q^2}_\tx{out}-\ol{\bs Q}_\tx{out}^2.
}
\item The time-like width-squared of the overlap of the in- and out-states:
\al{
\stin
	&=	{\sigma_\tx{in}\ov\Delta\bs V^2_\tx{in}},&
\stout
	&=	{\sigma_\tx{out}\ov\Delta\bs V^2_\tx{out}}.
}
\item The interaction time for the in- and out-states:
\al{
\mf T_\tx{in}
	&:=	{\ol{\bs V}_\tx{in}\cdot\ol{\bmf X}_\tx{in}
			-\ol{\bmf X\cdot\bs V}_\tx{in}
			\ov\Delta\bs V^2_\tx{in}},&
\mf T_\tx{out}
	&:=	{\ol{\bs V}_\tx{out}\cdot\ol{\bmf X}_\tx{out}
			-\ol{\bmf X\cdot\bs V}_\tx{out}
			\ov\Delta\bs V^2_\tx{out}},
			\label{interaction times}
}
\item The overlap exponent for the in- and out-states: 
\al{
\mc R_\tx{in}
	&:=	{\Delta\bmf X^2_\tx{in}\ov\sigma_\tx{in}}
		-{\mf T_\tx{in}^2\ov\varsigma_\tx{in}},&
\mc R_\tx{out}
	&:=	{\Delta\bmf X^2_\tx{out}\ov\sigma_\tx{out}}
		-{\mf T_\tx{out}^2\ov\varsigma_\tx{out}},
}
We can show the non-negativity of $\mc R_\tx{in}$ and $\mc R_\tx{out}$ as in Sec.~3.1 in Ref.~\cite{Ishikawa:2018koj}; our case corresponds to the $\sigma_0\to\infty$ limit in its Appendix~C.1.
\end{itemize}
We see from Eq.~\eqref{S master} that a configuration that has large $\mc R_\tx{in}$ or $\mc R_\tx{out}$ of initial and final-state phase space $\paren{\bs\Pi_1,\dots,\bs\Pi_4}$ and of the internal momentum $\bs p$ gives an exponentially suppressed wave-function overlap and the corresponding amplitude is also suppressed exponentially.

\subsection{Separation of bulk and time boundaries}
After integrating over $t$ and $t'$, we get
\al{
\mc S
	&=	i\kappa^2\paren{
			\prod_{A=1}^4{1\ov\sqrt{2E_A}}
			\paren{1\ov\pi\sigma_A}^{3/4}}
		\paren{2\pi\sigma_\tx{in}}^{3/2}
		\paren{2\pi\sigma_\tx{out}}^{3/2}
		\int{\df^4p\ov\paren{2\pi}^4}{1\ov p^2+M^2-i\epsilon}\nn
	&\quad\times
		\sqrt{2\pi\stin} \, G_\tx{in}\fn{\mc T_\tx{in}\fn{p}}\,
		\sqrt{2\pi\stout} \, G_\tx{out}\fn{\mc T_\tx{out}\fn{p}} \nn
	&\quad\times
		e^{-{\mc R_\tx{out}\ov2}}\,
		e^{
			-{\stout\ov2}\paren{p^0-\mc E_\tx{out}-\ol{\bs V}_\tx{out}\cdot\bs p}^2
			-i\mf T_\tx{out}\paren{p^0-\mc E_\tx{out}-\ol{\bs V}_\tx{out}\cdot\bs p}
			}\,
		e^{
			-{\sigma_\tx{out}\ov2}\paren{\bs p-\bs P_\tx{out}}^2
			+i\ol{\bmf X}_\tx{out}\cdot\paren{\bs p-\bs P_\tx{out}}
			}\nn		
	&\quad\times
		e^{-{\mc R_\tx{in}\ov2}}\,
		e^{
			-{\stin\ov2}\paren{p^0-\mc E_\tx{in}-\ol{\bs V}_\tx{in}\cdot\bs p}^2
			+i\mf T_\tx{in}\paren{p^0-\mc E_\tx{in}-\ol{\bs V}_\tx{in}\cdot\bs p}
			}\,
		e^{
			-{\sigma_\tx{in}\ov2}\paren{\bs p-\bs P_\tx{in}}^2
			-i\ol{\bmf X}_\tx{in}\cdot\paren{\bs p-\bs P_\tx{in}}
			},
			\label{exact result}
}
where
\al{
\mc E_\tx{in}
	&:=	E_\tx{in}-\ol{\bs V}_\tx{in}\cdot\bs P_\tx{in},\\
\mc E_\tx{out}
	&:=	E_\tx{out}-\ol{\bs V}_\tx{out}\cdot\bs P_\tx{out};
}
we have defined the window functions as in Ref.~\cite{Ishikawa:2018koj}
\al{
G_\tx{in}\fn{\mc T}
	&:=	\int_{T_\tx{in}}^{T_\tx{out}} \frac{\df t'}{\sqrt{2\pi\stin}}
		e^{
			-{1\ov2\stin}\paren{t' - \mc T}^2
			}, &
G_\tx{out}\fn{\mc T}
	&:=	\int_{T_\tx{in}}^{T_\tx{out}} \frac{\df t}{\sqrt{2\pi\stout}}
		e^{
			-{1\ov2\stout}\paren{t - \mc T}^2
			};
}
and
\al{
\mc T_\tx{in}\fn{p}
	&:=	\mf T_\tx{in} +i \stin\sqbr{
			\paren{p^0 - E_\tx{in}} 
			-\ol{\bs V}_\tx{in}\cdot \paren{\bs p - \bs P_\tx{in}}}\nn
	&=	\mf T_\tx{in} +i \stin\paren{
			p^0 - \mc E_\tx{in}-\ol{\bs V}_\tx{in}\cdot \bs p}, \nn
\mc T_\tx{out}\fn{p}
	&:=	\mf T_\tx{out}  -i \stout\sqbr{
			\paren{p^0 - E_\tx{out}} 
			-\ol{\bs V}_\tx{out}\cdot \paren{\bs p - \bs P_\tx{out}}}\nn
	&=	\mf T_\tx{out}  -i \stout\paren{
			p^0 - \mc E_\tx{out}
			-\ol{\bs V}_\tx{out}\cdot\bs p}.
}
Physically, the complex variable $\mc T_\tx{in}$ ($\mc T_\tx{out}$), or especially its real part $\Re\mc T_\tx{in}=\mf T_\tx{in}$ ($\Re\mc T_\tx{out}=\mf T_\tx{out}$), corresponds to an ``interaction time'' at which the interaction occurs between the initial (final) $\phi\phi$ and the internal $\Phi$.

In terms of the Gauss error function
\al{
\erf\fn{z} := {2\ov\sqrt{\pi}} \int_{0}^{z} e^{-x^2} dx,
}
the above two functions are represented as follows:
\al{
G_\tx{in-int}(\mc T)
	&=	{1\ov2} 
		\left[
		\erf\fn{  {\mc T  - T_\tx{in} }\ov{\sqrt{2 \stin}}  } -
		\erf\fn{  {\mc T  - T_\tx{out} }\ov{\sqrt{2 \stin}}  }
		\right],\nn
G_\tx{out-int}(\mc T)
	&=	{1\ov2} 
		\left[
		\erf\fn{  {\mc T  - T_\tx{in} }\ov{\sqrt{2 \stout}}  } -
		\erf\fn{  {\mc T  - T_\tx{out} }\ov{\sqrt{2 \stout}}  }
		\right].
}
For convenience, we distinguish the bulk effects from the in- and out-boundary ones as
\al{
G_\tx{in-int}(\mc T)
	&:= G_\tx{in-int}^\text{bulk}(\mc T) + G_\tx{in-int}^\text{in-bdry}(\mc T) + G_\tx{in-int}^\text{out-bdry}(\mc T),\nn
G_\tx{out-int}(\mc T)
	&:= G_\tx{out-int}^\text{bulk}(\mc T) + G_\tx{out-int}^\text{in-bdry}(\mc T) + G_\tx{out-int}^\text{out-bdry}(\mc T),
}
where for the interaction between the initial $\phi\phi$ state and the intermediate $\Phi$,
\al{
G_\tx{in-int}^\text{bulk}(\mc T)
	&:=	{1\ov2} 
		\left[
		\sgn\fn{  {\mc T  - T_\tx{in} }\ov{\sqrt{2 \stin}}  } -
		\sgn\fn{  {\mc T  - T_\tx{out} }\ov{\sqrt{2 \stin}}  }
		\right],\nn
G_\tx{in-int}^\text{in-bdry}(\mc T)
	&:=	{1\ov2} 
		\left[
		\erf\fn{  {\mc T  - T_\tx{in} }\ov{\sqrt{2 \stin}}  } -
		\sgn\fn{  {\mc T  - T_\tx{in} }\ov{\sqrt{2 \stin}}  }
		\right],\nn
G_\tx{in-int}^\text{out-bdry}(\mc T)
	&:=	{1\ov2} 
		\left[
		\sgn\fn{  {\mc T  - T_\tx{out} }\ov{\sqrt{2 \stin}}  } -
		\erf\fn{  {\mc T  - T_\tx{out} }\ov{\sqrt{2 \stin}}  }
		\right],
}
and for the interaction between the final $\phi\phi$ state and the intermediate $\Phi$,
\al{
G_\tx{out-int}^\text{bulk}(\mc T)
	&:=	{1\ov2} 
		\left[
		\sgn\fn{  {\mc T  - T_\tx{in} }\ov{\sqrt{2 \stout}}  } -
		\sgn\fn{  {\mc T  - T_\tx{out} }\ov{\sqrt{2 \stout}}  }
		\right],\nn
G_\tx{out-int}^\text{in-bdry}(\mc T)
	&:=	{1\ov2} 
		\left[
		\erf\fn{  {\mc T  - T_\tx{in} }\ov{\sqrt{2 \stout}}  } -
		\sgn\fn{  {\mc T  - T_\tx{in} }\ov{\sqrt{2 \stout}}  }
		\right],\nn
G_\tx{out-int}^\text{out-bdry}(\mc T)
	&:=	{1\ov2} 
		\left[
		\sgn\fn{  {\mc T  - T_\tx{out} }\ov{\sqrt{2 \stout}}  } -
		\erf\fn{  {\mc T  - T_\tx{out} }\ov{\sqrt{2 \stout}}  }
		\right].
}
Here, the following sign function for a complex variable has been defined:
\al{
\sgn\fn{z}
	&:=	\begin{cases}
		+1 & \tx{for } \Re z > 0 \tx{ or } \paren{\Re z = 0 \tx{ and } \Im z > 0}, \\
		-1 & \tx{for } \Re z < 0 \tx{ or } \paren{\Re z = 0 \tx{ and } \Im z < 0}, \\
		0  & \tx{for } z = 0.
		\end{cases}
}
More explicitly,
\al{
G_\tx{in-int}^\text{bulk}(\mc T)
	&=	\begin{cases}
		1 & \paren{T_\tx{in} < \Re\mc T < T_\tx{out}}, \\
		0 & \paren{\Re\mc T < T_\tx{in} \text{ or } T_\tx{out} < \Re\mc T}, \\
		\theta\fn{+{\Im\mc T\ov\stin}} & \paren{\Re\mc T = T_\tx{in}}, \\
		\theta\fn{-{\Im\mc T\ov\stin}} & \paren{\Re\mc T = T_\tx{out}},
		\end{cases}\nn
G_\tx{out-int}^\text{bulk}(\mc T)
	&=	\begin{cases}
		1 & \paren{T_\tx{in} < \Re\mc T < T_\tx{out}}, \\
		0 & \paren{\Re\mc T < T_\tx{in} \text{ or } T_\tx{out} < \Re\mc T}, \\
		\theta\fn{+{\Im\mc T\ov\stout}} & \paren{\Re\mc T = T_\tx{in}}, \\
		\theta\fn{-{\Im\mc T\ov\stout}} & \paren{\Re\mc T = T_\tx{out}},
		\end{cases}
}
where we define the step function for a real variable as
\al{
\theta(x)
	= 	{{1 + \sgn\fn{x}}\ov2}
	=	\begin{cases}
		1 & \paren{x > 0}, \\
		1\ov2 & \paren{x = 0}, \\
		0 & \paren{x < 0}. \\
		\end{cases}
}
Detailed discussion for the boundary terms can be found in Ref.~\cite{Ishikawa:2018koj}.

Under the above classification of the in- and  out-window functions,
we divide the probability amplitude $\mc S$ into two parts:
\al{
\mc S = \mc S_\tx{bulk} + \mc S_\tx{boundary},
}
where $S_\tx{bulk}$ contains the pure bulk contributions from $G_\tx{in-int}^\text{bulk}(\mc T_\tx{in})$ and
$G_\tx{out-int}^\text{bulk}(\mc T_\tx{out})$, while every term of $\mc S_\tx{boundary}$ includes at least
one boundary window function.

\section{Interpretation of boundary effect}\label{boundary effect section}

\begin{figure}\centering
\includegraphics[width=\textwidth]{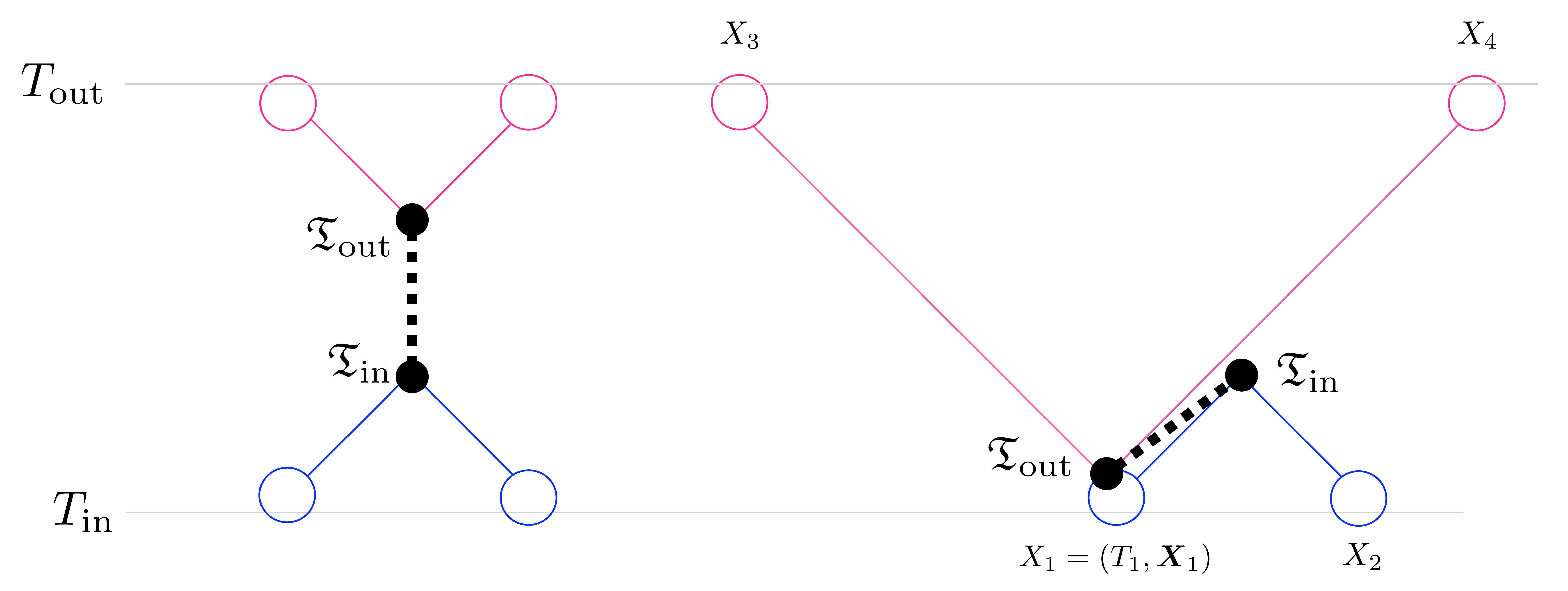}
\caption{\label{schematic figure}
Schematic diagram in position space. Each of blue and red lines denotes the trajectory of the center of wave-packet for in and out states $\phi$, respectively. The thick dashed line denotes the trajectory of internal particle $\Phi$, while the black dots of its ends indicate that the interactions occur in a finite range with the spatial and time-like widths $\sim\sqrt{\sigma_\tx{in}}$ and $\sqrt{\stin}$ ($\sim\sqrt{\sigma_\tx{out}}$ and $\sqrt{\stout}$) around the point $\ol{\bs\Xi\fn{\mf T_\tx{in}}}_\tx{in}$ at time $\mf T_\tx{in}$ (point $\ol{\bs\Xi\fn{\mf T_\tx{out}}}_\tx{out}$ at time $\mf T_\tx{out}$), respectively. Circles are a reminder that each packet is given with a finite width, namely with the widths $\sim\sqrt{\sigma_1}$ and $\sqrt{\sigma_2}$ ($\sim\sqrt{\sigma_3}$ and $\sqrt{\sigma_4}$) at times $T_1$ and $T_2$ ($T_3$ and $T_4$) for the initial (final) wave packets. In the perturbation theory, we consider time evolution of the in-state from $T_\tx{in}$ to $T_\tx{out}$ in the interaction picture, which are chosen near $T_1,T_2$ and $T_3,T_4$, respectively, and the S-matrix element is taken with the out-state at $T_\tx{out}$.
The left figure shows an $s$-channel scattering without a backward propagation in the sense of the old-fashioned perturbation theory.
The right figure explains that there always exists a final state configuration that realizes, e.g.\ $T_1\simeq\mf T_\tx{out}$ no matter how large we take a cluster-decomposition limit: $\ab{\bs\Xi_1\fn{T_\tx{in}}-\bs\Xi_2\fn{T_\tx{in}}}\sim\ab{\bs X_1-\bs X_2}\to\infty$.}
\end{figure}

We present and clarify two different interpretations of the result~\eqref{exact result}.
We consider a finite time interval $T_\tx{out}-T_\tx{in}$.
Without loss of generality, we focus on the initial time boundary at $T_\tx{in}$ unless otherwise stated.
First we stress that when we integrate over the final-state phase space $\bs\Pi_3$ and $\bs\Pi_4$ with varying interaction time $\mf T_\tx{out}$ ($=\Re\mc T_\tx{out}$) accordingly to Eq.~\eqref{interaction times}, there always exists a final-state configuration that gives a significant in-boundary effect at $T_\tx{in}$, no matter what initial configuration we take, even a cluster-decomposition limit $\ab{\bs\Xi_1\fn{T_\tx{in}}-\bs\Xi_2\fn{T_\tx{in}}}\to\infty$ and/or take $T_\tx{in}\to-\infty$; see Fig.~\ref{schematic figure}.

To illustrate qualitative behavior, let us tentatively focus on the expressions in the following limit~\cite{Ishikawa:2018koj}:\footnote{
Hereafter we sometimes use $\mc T$ for $\mc T_\tx{out}$ just for presentation.
More precisely, we should rather write $\mc T_\tx{out-int}$ and $T_\tx{in-bdry}$, but this would be too cumbersome.
}
\al{
\ab{\mc T-T_\tx{in}}
	&\gg	{\sqrt{2 \stout}},
		\label{illustrative limit}
}
which results in\footnote{
In Eq.~\eqref{illustrative limit of G}, we cannot take ${\ab{\mc T_\tx{out}-T_\tx{in}}\ov\sqrt{2\stout}}\to0$ limit because of the assumption~\eqref{illustrative limit}. When correctly taken, this limit is finite; see Ref.~\cite{Ishikawa:2018koj}.
}
\al{
G_\tx{out-int}\fn{\mc T}
	&\to	G_\tx{out-int}^\tx{bulk}\fn{\mc T}
			-{1\ov\sqrt\pi}e^{-{\paren{\mc T-T_\tx{in}}^2\ov2\stout}}{\sqrt{2\stout}\ov\mc T-T_\tx{in}}.
			\label{illustrative limit of G}
}
Note that the illustrative limit~\eqref{illustrative limit} implies that near the boundary, $\paren{\Re\mc T_\tx{out}-T_\tx{in}}^2\lesssim2\stout$, the deviation from the ``energy conservation'' is large:
\al{
\paren{p^0-\mc E_\tx{out}-\ol{\bs V}_\tx{out}\cdot\bs p}^2=\paren{\Im\mc T_\tx{out}}^2\gg2\stout.
}
From Eq.~\eqref{illustrative limit of G}, we see that the boundary effect may become significant when $\mc T$ is near the in-boundary, namely when $\paren{\Re\mc T-T_\tx{in}}^2\lesssim2\stout$ with $\paren{\Im\mc T}\gg2\stout$ as said above:
\al{
G_\tx{out-int}\fn{\mc T}
	&\to
		-{1\ov\sqrt\pi}e^{{\paren{\Im\mc T}^2\ov2\stout}}{\sqrt{2\stout}\ov i\,\Im\mc T}.
}
Note that the apparent exponential growth for the energy non-conserving limit $\paren{\Im\mc T}^2\gg2\stout$ is cancelled out by the existing energy conservation factor coming from
\al{
e^{-{\stout\ov2}\paren{p^0-\mc E_\tx{out}-\ol{\bs V}_\tx{out}\cdot\bs p}^2}
	&=	e^{-{\paren{\Im\mc T_\tx{out}}^2\ov2\stout}}.
}
That is, the exponential suppression factor for a deviation from the energy conservation, $e^{-{\paren{\Im\mc T_\tx{out}}^2/2\stout}}$, is cencelled and replaced by the power suppression factor $1/\Im\mc T$ in the boundary effect.
Recall that the boundary contribution from the configuration $\paren{\Re {\mc T}_\tx{out}-\mf T_\tx{in}}^2\lesssim2 \varsigma_\tx{out}$ arises even if $\bs X_3$ and $\bs X_4$ are at a distance.\footnote{
Suppose we consider the probability from the amplitude~\eqref{exact result}, $P=\ab{\mc S}^2$, for a special case $T_1=T_2=T_\tx{in}$ and $T_3=T_4=T_\tx{out}$: $P\fn{T_\tx{in}\bs\Pi_1\bs\Pi_2\to T_\tx{out}\bs\Pi_3\bs\Pi_4}$.
It satisfies $P\fn{T_\tx{in}\bs\Pi_1\bs\Pi_2\to T_\tx{out}\bs\Pi_3\bs\Pi_4}\to0$ in the limits $T_\tx{out}\to T_\tx{in}$ and $\ab{\bs X_i-\bs X_j}\to\infty$ for all $i=1,2$ and $j=3,4$. We also have $P\fn{T_\tx{in}\bs\Pi_1\bs\Pi_2\to T_\tx{in}\bs\Pi_1\bs\Pi_2}=1$. Here,  $P\fn{T_\tx{in}\bs\Pi_1\bs\Pi_2 \rightarrow T_\tx{out}\bs\Pi_3\bs\Pi_4}$ represents a transition probability for not only short distance interactions but also long distance ones such as the Coulomb potential; see also the discussion below Eq.~\eqref{Dyson series}.
}

The existence of boundary effect crucially depends on the relation~\eqref{boundary relation}.
The key question is the following:
Can we well approximate the real physical setup in experiment, namely the Schr\"odinger-picture in-state $e^{-i\h Ht}\Ket{\tx{in; }\Pi_1\Pi_2}$, by the ``free Schr\"odinger-picture'' state $e^{-i\h H_\tx{free}t}\Ket{\Pi_1\Pi_2}$, evolving in a virtual free world without any interaction, at $t=T_\tx{in}$ when interactions are not negligible?\footnote{
In this section, we omit to show the trivial dependence on $\sigma_1$, $\sigma_2$, etc.
}
If not, what state should we prepare for $e^{-i\h Ht}\Ket{\tx{in; }\Pi_1\Pi_2}$ at $t=T_\tx{in}$?
Here we introduce two different constructions: ``free'' and ``dressed'', which say yes and no for the first question, respectively.

\subsection{Quantum mechanics basics}
For the discussion below, let us recall the basics of quantum mechanics and spell out our notation. We identify the Schr\"odinger, Heisenberg, and interaction pictures at an arbitrary reference time $t_\tx{r}$: For an arbitrary operator $\h O$ in the Schr\"odinger picture, we relate them by\footnote{
Recall that in the interaction picture, we separate an expectation value as
\als{
\,\paren{\Bra{{\sf\Psi}}e^{i\h H\,\paren{t-t_\tx{r}}}e^{-i\h H_\tx{free}\,\paren{t-t_\tx{r}}}}
\,\paren{e^{i\h H_\tx{free}\,\paren{t-t_\tx{r}}}\h Oe^{-i\h H_\tx{free}\,\paren{t-t_\tx{r}}}}
\,\paren{e^{i\h H_\tx{free}\,\paren{t-t_\tx{r}}}e^{-i\h H\,\paren{t-t_\tx{r}}}\Ket{{\sf\Psi}}}.
}
}
\al{
\h O^\tx{I}\fn{t}
	&=	e^{i\h H_\tx{free}\,\paren{t-t_\tx{r}}}\h Oe^{-i\h H_\tx{free}\,\paren{t-t_\tx{r}}},\\
\h O^\tx{H}\fn{t}
	&=	e^{i\h H\,\paren{t-t_\tx{r}}}\h Oe^{-i\h H\,\paren{t-t_\tx{r}}},
}
and for a time-independent state $\Ket{{\sf\Psi}}$ in the Heisenberg picture by
\al{
\Ket{{\sf\Psi};t}_\tx{I}
	&=	e^{i\h H_\tx{free}\,\paren{t-t_\tx{r}}}e^{-i\h H\,\paren{t-t_\tx{r}}}\Ket{{\sf\Psi}}\nn
	&=	\paren{
			\T e^{-i\int_{t_\tx{r}}^t\h H_\tx{int}^\tx{I}\fn{t'-t_\tx{r}}\,\df t'}
			}
		\Ket{{\sf\Psi}},
		\label{interaction picture state}\\
\Ket{{\sf\Psi};t}_\tx{S}
	&=	e^{-i\h H\,\paren{t-t_\tx{r}}}\Ket{{\sf\Psi}},
		\label{Schroedinger picture state}
}
where we have used
\al{
\h U\fn{t_1,t_2}
	&:=	e^{i\h H_\tx{free}\,\paren{t_1-t_\tx{r}}}e^{-i\h H\,\paren{t_1-t_2}}e^{-i\h H_\tx{free}\,\paren{t_2-t_\tx{r}}}\nn
	&=	\T e^{-i\int_{t_2-t_\tx{r}}^{t_1-t_\tx{r}}\h H_\tx{int}^\tx{I}\fn{t'}\,\df t'}
	=	\T e^{-i\int_{t_2}^{t_1}\h H_\tx{int}^\tx{I}\fn{t'-t_\tx{r}}\,\df t'}.
}
If an eigenbasis $\Ket{{\sf\Phi}}$ exist in the Schr\"odinger picture, $\h O\Ket{{\sf\Phi}}=o\Ket{{\sf\Phi}}$, the corresponding operators in the interaction and Heisenberg pictures have the following eigenbases, respectively:
\al{
\Ket{{\sf\Phi};t}_\tx{IB}
	&=	e^{i\h H_\tx{free}\,\paren{t-t_\tx{r}}}\Ket{{\sf\Phi}},\\
\Ket{{\sf\Phi};t}_\tx{HB}
	&=	e^{i\h H\,\paren{t-t_\tx{r}}}\Ket{{\sf\Phi}}.
}
The time dependence of these eigenbases is different from that of the states~\eqref{interaction picture state} and \eqref{Schroedinger picture state}. Typically in our computation, $\h O$ stands for $\h H_\tx{free}$.

\subsection{``Free'' construction}
So far, we have chosen an arbitrary initial (final) time $T_\tx{in}$ ($T_\tx{out}$) anywhere near $T_1$ ($T_3$) and/or $T_2$ ($T_4$).
In the ``free'' construction we identify the in and out Schr\"odinger-picture states at times $T_\tx{in}$ and $T_\tx{out}$, respectively, with a ``free Schr\"odinger picture'' state that evolves in a virtual free world governed by the free Hamiltonian no matter how significant interactions are at these times:
\al{
\Ket{\tx{in; }\Pi_1\Pi_2;t=T_\tx{in}}_\tx{S}
	&=	\Ket{\Pi_1\Pi_2;t=T_\tx{in}}_\tx{S}^\tx{free},
	\label{identification again}\nn
\Ket{\tx{out; }\Pi_3\Pi_4;t=T_\tx{out}}_\tx{S}
	&=	\Ket{\Pi_3\Pi_4;t=T_\tx{out}}_\tx{S}^\tx{free},
}
where we have defined the ``free Schr\"odinger'' state that evolves in the virtual free world:
\al{
\Ket{{\sf\Psi};t}_\tx{S}^\tx{free}
	&:=	e^{-i\h H_\tx{free}\,\paren{t-t_\tx{r}}}\Ket{{\sf\Psi}}.
}
In other words, the in and out states are given in the Heisenberg picture as
\al{
\Ket{\tx{in; }\Pi_1\Pi_2}
	&=	e^{i\h H\,\paren{T_\tx{in}-t_\tx{r}}}e^{-i\h H_\tx{free}\,\paren{T_\tx{in}-t_\tx{r}}}\Ket{\Pi_1\Pi_2},\nn
\Ket{\tx{out; }\Pi_3\Pi_4}
	&=	e^{i\h H\,\paren{T_\tx{out}-t_\tx{r}}}e^{-i\h H_\tx{free}\,\paren{T_\tx{out}-t_\tx{r}}}\Ket{\Pi_3\Pi_4};
		\label{Heisenberg picture relation}
}
in the Schr\"odinger picture as
\al{
\Ket{\tx{in; }\Pi_1\Pi_2;t}_\tx{S}
	&=	e^{-i\h H\,\paren{t-t_\tx{r}}}\paren{e^{i\h H\,\paren{T_\tx{in}-t_\tx{r}}}e^{-i\h H_\tx{free}\,\paren{T_\tx{in}-t_\tx{r}}}\Ket{\Pi_1\Pi_2}}\nn
	&=	e^{-i\h H\,\paren{t-T_\tx{in}}}e^{-i\h H_\tx{free}\,\paren{T_\tx{in}-t_\tx{r}}}\Ket{\Pi_1\Pi_2},\nn
\Ket{\tx{out; }\Pi_3\Pi_4;t}_\tx{S}
	&=	e^{-i\h H\,\paren{t-t_\tx{r}}}\paren{e^{i\h H\,\paren{T_\tx{out}-t_\tx{r}}}e^{-i\h H_\tx{free}\,\paren{T_\tx{out}-t_\tx{r}}}\Ket{\Pi_3\Pi_4}}\nn
	&=	e^{-i\h H\,\paren{t-T_\tx{out}}}e^{-i\h H_\tx{free}\,\paren{T_\tx{out}-t_\tx{r}}}\Ket{\Pi_3\Pi_4};
}
and in the interaction picture as
\al{
\Ket{\tx{in; }\Pi_1\Pi_2;t}_\tx{I}
	&=	e^{i\h H_\tx{free}\,\paren{t-t_\tx{r}}}e^{-i\h H\,\paren{t-t_\tx{r}}}
		\paren{e^{i\h H\,\paren{T_\tx{in}-t_\tx{r}}}e^{-i\h H_\tx{free}\,\paren{T_\tx{in}-t_\tx{r}}}\Ket{\Pi_1\Pi_2}}\nn
	&=	e^{i\h H_\tx{free}\,\paren{t-t_\tx{r}}}
		e^{-i\h H\,\paren{t-T_\tx{in}}}
		e^{-i\h H_\tx{free}\,\paren{T_\tx{in}-t_\tx{r}}}\Ket{\Pi_1\Pi_2}\nn
	&=	\T e^{-i\int_{T_\tx{in}}^t\h H_\tx{int}^\tx{I}\fn{t'-t_\tx{r}}\,\df t'}
		\Ket{\Pi_1\Pi_2},\nn
\Ket{\tx{out; }\Pi_3\Pi_4;t}_\tx{I}
	&=	e^{i\h H_\tx{free}\,\paren{t-t_\tx{r}}}e^{-i\h H\,\paren{t-t_\tx{r}}}
		\paren{e^{i\h H\,\paren{T_\tx{out}-t_\tx{r}}}e^{-i\h H_\tx{free}\,\paren{T_\tx{out}-t_\tx{r}}}\Ket{\Pi_3\Pi_4}}\nn
	&=	e^{i\h H_\tx{free}\,\paren{t-t_\tx{r}}}
		e^{-i\h H\,\paren{t-T_\tx{out}}}
		e^{-i\h H_\tx{free}\,\paren{T_\tx{out}-t_\tx{r}}}\Ket{\Pi_3\Pi_4}\nn
	&=	\T e^{-i\int_{T_\tx{out}}^t\h H_\tx{int}^\tx{I}\fn{t'-t_\tx{r}}\,\df t'}
		\Ket{\Pi_3\Pi_4}.
}
One can trivially check the following:
\al{
\Ket{\tx{in; }\Pi_1\Pi_2}
	&=	\Ket{\tx{in; }\Pi_1\Pi_2;t_\tx{r}}_\tx{S}
	=	\Ket{\tx{in; }\Pi_1\Pi_2;t_\tx{r}}_\tx{I},\nn
\Ket{\tx{out; }\Pi_3\Pi_4}
	&=	\Ket{\tx{out; }\Pi_3\Pi_4;t_\tx{r}}_\tx{S}
	=	\Ket{\tx{out; }\Pi_3\Pi_4;t_\tx{r}}_\tx{I}.
}
We also see that the Heisenberg-picture relation~\eqref{Heisenberg picture relation} reads in the Schr\"odinger picture,
\al{
\Ket{\tx{in; }\Pi_1\Pi_2;T_\tx{in}}_\tx{S}
	&=	e^{-i\h H_\tx{free}\,\paren{T_\tx{in}-t_\tx{r}}}\Ket{\Pi_1\Pi_2},\nn
\Ket{\tx{out; }\Pi_3\Pi_4;T_\tx{out}}_\tx{S}
	&=	e^{-i\h H_\tx{free}\,\paren{T_\tx{out}-t_\tx{r}}}\Ket{\Pi_3\Pi_4},
}
and in the interaction picture,
\al{
\Ket{\tx{in; }\Pi_1\Pi_2;T_\tx{in}}_\tx{I}
	&=	\Ket{\Pi_1\Pi_2},\nn
\Ket{\tx{out; }\Pi_3\Pi_4;T_\tx{out}}_\tx{I}
	&=	\Ket{\Pi_3\Pi_4}.
		\label{interaction picture identification}
}

The ``free'' construction puts more emphasis on the interaction picture, in which the identification~\eqref{interaction picture identification} appears most natural. We can also rewrite the probability amplitude as an inner product of the interaction-picture states at an arbitrary time~$t$:
\al{
\mc S
	&=	{}_\tx{I}\!\Braket{\tx{out; }\Pi_3\Pi_4; t|\tx{in; }\Pi_1\Pi_2; t}_\tx{I}\nn
	&=	\Bra{\Pi_3\Pi_4}\T e^{i\int^t_{T_\tx{out}}\h H_\tx{int}^\tx{I}\fn{t'-t_\tx{r}}\,\df t'}e^{-i\int^t_{T_\tx{in}}\h H_\tx{int}^\tx{I}\fn{t'-t_\tx{r}}\,\df t'}\Ket{\Pi_1\Pi_2}\nn
	&=	\Bra{\Pi_3\Pi_4}\T e^{-i\int^{T_\tx{out}}_{T_\tx{in}}\h H_\tx{int}^\tx{I}\fn{t'-t_\tx{r}}\,\df t'}\Ket{\Pi_1\Pi_2},
		\label{S-matrix in free construction}
}
which becomes Eq.~\eqref{Dyson series} when we set the arbitrary reference time  $t_\tx{r}=0$ as before.\footnote{
Or else, we may rewrite
\als{
\mc S
	&=	\paren{\Bra{\Pi_3\Pi_4}e^{-i\h H_\tx{free}t_\tx{r}}}
		\paren{\T e^{-i\int^{T_\tx{out}}_{T_\tx{in}}\h H_\tx{int}^\tx{I}\fn{t'}\,\df t'}}
		\paren{e^{i\h H_\tx{free}t_\tx{r}}\Ket{\Pi_1\Pi_2}},
}
and redefine all the free states $e^{i\h H_\tx{free}t_\tx{r}}\Ket{\Phi}$, each being an $\h H_\tx{free}$-eigenstate, to be $\Ket{\Phi}$.
}
Note that the $t$ dependence drops out of the expression, and hence the probability does not depend on~$t$.

We may say that the boundary effects remain even if the interaction is taken into account in the following sense~\cite{Ishikawa:2016lnn} (see also Ref.~\cite{Kamefuchi:1961sb}):
Suppose that we transform the free states by a unitary operator $\h V\fn{\kappa}$ with $\h V^\dagger\fn{\kappa}\h V\fn{\kappa}=\h1$ in Eq.~\eqref{S-matrix in free construction}:
\al{
\wt{\Ket{\Pi_1\Pi_2}}
	&=	\h V\fn{\kappa}\Ket{\Pi_1\Pi_2},\\
\wt{\Ket{\Pi_3\Pi_4}}
	&=	\h V\fn{\kappa}\Ket{\Pi_3\Pi_4}.
}
Then the S-matrix becomes
\al{
\wt{\mc S}
	&=	\wt{\Bra{\Pi_3\Pi_4}}\h U\fn{T_\tx{out},T_\tx{in}}\wt{\Ket{\Pi_1\Pi_2}}\nn
	&=	\Bra{\Pi_3\Pi_4}
		\h V^\dagger\fn{\kappa}
		\h U\fn{T_\tx{out},T_\tx{in}}
		\h V\fn{\kappa}
		\Ket{\Pi_1\Pi_2}.
}
If $\h V$ is expanded as $\h V=\h1+\Or{\kappa}$, we see from $\h U\fn{T_\tx{out},T_\tx{in}}=\h1+\Or{\kappa^2}$ that
\al{
\commutator{\h V\fn{\kappa}}{\h U\fn{T_\tx{out},T_\tx{in}}}
	&=	\Or{\kappa^3},
}
and hence
\al{
\h V^\dagger\fn{\kappa}
		\h U\fn{T_\tx{out},T_\tx{in}}
		\h V\fn{\kappa}
	&=	\h U\fn{T_\tx{out},T_\tx{in}}
		+\Or{\kappa^3}.
}
Accordingly the order $\kappa^2$ contribution of the transition amplitudes are  invariant under the unitary change of the free states.

\subsection{``Dressed'' construction}\label{dressed section}
To repeat, we have chosen an arbitrary initial time $T_\tx{in}$ anywhere near $T_1$ and/or $T_2$.
One might feel it strange to identify the initial state as in Eq.~\eqref{identification again} for a wave-packet configuration $\paren{\bs\Pi_1,\dots,\bs\Pi_4}$ that gives a significant overlap of the final-state wave-packets at $\mf T_\tx{out}\simeq T_\tx{in}$ so that interactions are not negligible at $T_\tx{in}$ as in the right panel in Fig.~\ref{schematic figure}.
In particular, the boundary interaction~\eqref{illustrative limit of G} crucially depends on the arbitrarily chosen $T_\tx{in}$: For a given fixed initial and final state configuration $\paren{\bs\Pi_1,\dots,\bs\Pi_4}$, the boundary contribution drops off exponentially as we shift the arbitrarily chosen $T_\tx{in}$ backwards in time.

The boundary effect is a consequence of the above-mentioned identification of the Heisenberg state $\Ket{\tx{in; }\Pi_1\Pi_2}$ and $\Ket{\tx{out; }\Pi_3\Pi_4}$ at $T_\tx{in}$ and $T_\tx{out}$, respectively.
What if we identify different states at $T_\tx{in}$ and $T_\tx{out}$?
Suppose that we take into account the interactions from $T_\tx{in}'$ $(<T_\tx{in})$ to $T_\tx{in}$ and from $T_\tx{out}'$ ($>T_\tx{out}$) to $T_\tx{out}$ (backward in time as $T_\tx{out}<T_\tx{out}'$) in addition to the ``free'' construction above:
\al{
\Ket{\tx{in; }\Pi_1\Pi_2}'
	&=	e^{i\h H\,\paren{T_\tx{in}-t_\tx{r}}}e^{-i\h H_\tx{free}\,\paren{T_\tx{in}-t_\tx{r}}}\T e^{-i\int_{T_\tx{in}'}^{T_\tx{in}}\h H_\tx{int}\fn{t'-t_\tx{r}}\,\df t'}\Ket{\Pi_1\Pi_2},
		\nn
\Ket{\tx{out; }\Pi_3\Pi_4}'
	&=	e^{i\h H\,\paren{T_\tx{out}-t_\tx{r}}}e^{-i\h H_\tx{free}\,\paren{T_\tx{out}-t_\tx{r}}}\T e^{-i\int^{T_\tx{out}}_{T_\tx{out}'}\h H_\tx{int}\fn{t'-t_\tx{r}}\,\df t'}\Ket{\Pi_3\Pi_4},
	\label{dressed Heisenberg states}
}
where we have replaced $\Ket{\Pi_1\Pi_2}$ and $\Ket{\Pi_3\Pi_4}$ in the ``free'' construction~\eqref{Heisenberg picture relation} by
\al{
\Ket{\Pi_1\Pi_2}
	&\to	\T e^{-i\int_{T_\tx{in}'}^{T_\tx{in}}\h H_\tx{int}\fn{t'-t_\tx{r}}\,\df t'}\Ket{\Pi_1\Pi_2},
		\nn
\Ket{\Pi_3\Pi_4}
	&\to	\T e^{-i\int^{T_\tx{out}}_{T_\tx{out}'}\h H_\tx{int}\fn{t'-t_\tx{r}}\,\df t'}\Ket{\Pi_3\Pi_4}.
		\label{replacement of states}
}
We note that the free basis $\Ket{\Pi_1\Pi_2}$ and the state $\T e^{-i\int_{T_\tx{in}'}^{T_\tx{in}}\h H_\tx{int}\fn{t'-t_\tx{r}}\,\df t'}\Ket{\Pi_1\Pi_2}$ are different from each other; the same note applies for the out ones.
Note also that we can rewrite the Heisenberg-picture states~\eqref{dressed Heisenberg states} as
\al{
\Ket{\tx{in; }\Pi_1\Pi_2}'
	&=	e^{i\h H\,\paren{T_\tx{in}-t_\tx{r}}}e^{-i\h H_\tx{free}\,\paren{T_\tx{in}-t_\tx{r}}}\nn
	&\quad \times
		\paren{
			e^{i\h H_\tx{free}\,\paren{T_\tx{in}-t_\tx{r}}}e^{-i\h H\,\paren{T_\tx{in}-t_\tx{r}}}
			e^{i\h H\paren{T_\tx{in}'-t_\tx{r}}}e^{-i\h H_\tx{free}\paren{T_\tx{in}'-t_\tx{r}}}
			}
		\Ket{\Pi_1\Pi_2}\nn
	&=	e^{i\h H\paren{T_\tx{in}'-t_\tx{r}}}e^{-i\h H_\tx{free}\paren{T_\tx{in}'-t_\tx{r}}}
		\Ket{\Pi_1\Pi_2},\nn
\Ket{\tx{out; }\Pi_3\Pi_4}'
	&=	e^{i\h H\,\paren{T_\tx{out}-t_\tx{r}}}e^{-i\h H_\tx{free}\,\paren{T_\tx{out}-t_\tx{r}}}\nn
	&\quad \times
		\paren{
			e^{i\h H_\tx{free}\,\paren{T_\tx{out}-t_\tx{r}}}e^{-i\h H\,\paren{T_\tx{out}-t_\tx{r}}}
			e^{i\h H\paren{t_\tx{out}'-t_\tx{r}}}e^{-i\h H_\tx{free}\paren{T_\tx{out}'-t_\tx{r}}}
			}
		\Ket{\Pi_1\Pi_2}\nn
	&=	e^{i\h H\paren{T_\tx{out}'-t_\tx{r}}}e^{-i\h H_\tx{free}\paren{T_\tx{out}'-t_\tx{r}}}
		\Ket{\Pi_3\Pi_4}.
		\label{dressed states}
}
In the Schr\"odinger picture, these are equivalent to
\al{
\Ket{\tx{in; }\Pi_1\Pi_2;t}'_\tx{S}
	&=	e^{-i\h H\,\paren{t-t_\tx{r}}}\paren{e^{i\h H\paren{T_\tx{in}'-t_\tx{r}}}e^{-i\h H_\tx{free}\paren{T_\tx{in}'-t_\tx{r}}}
		\Ket{\Pi_1\Pi_2}}\nn
	&=	e^{-i\h H\,\paren{t-T_\tx{in}'}}e^{-i\h H_\tx{free}\,\paren{T_\tx{in}'-t_\tx{r}}}\Ket{\Pi_1\Pi_2},\nn
\Ket{\tx{out; }\Pi_3\Pi_4;t}'_\tx{S}
	&=	e^{-i\h H\,\paren{t-t_\tx{r}}}\paren{e^{i\h H\paren{T_\tx{out}'-t_\tx{r}}}e^{-i\h H_\tx{free}\paren{T_\tx{out}'-t_\tx{r}}}
		\Ket{\Pi_3\Pi_4}}\nn
	&=	e^{-i\h H\,\paren{t-T_\tx{out}'}}e^{-i\h H_\tx{free}\,\paren{T_\tx{out}'-t_\tx{r}}}\Ket{\Pi_3\Pi_4},
}
and in the interaction picture,
\al{
\Ket{\tx{in; }\Pi_1\Pi_2;t}'_\tx{I}
	&=	e^{i\h H_\tx{free}\,\paren{t-t_\tx{r}}}e^{-i\h H\,\paren{t-t_\tx{r}}}
		\paren{e^{i\h H\paren{T_\tx{in}'-t_\tx{r}}}e^{-i\h H_\tx{free}\paren{T_\tx{in}'-t_\tx{r}}}\Ket{\Pi_1\Pi_2}}\nn
	&=	e^{i\h H_\tx{free}\,\paren{t-t_\tx{r}}}e^{-i\h H\,\paren{t-T_\tx{in}'}}
		e^{-i\h H_\tx{free}\paren{T_\tx{in}'-t_\tx{r}}}\Ket{\Pi_1\Pi_2}\nn
	&=	\T e^{-i\int_{T_\tx{in}'}^t\h H^\tx{I}_\tx{int}\fn{t'-t_\tx{r}}\,\df t'}\Ket{\Pi_1\Pi_2},\\
\Ket{\tx{out; }\Pi_3\Pi_3;t}'_\tx{I}
	&=	e^{i\h H_\tx{free}\,\paren{t-t_\tx{r}}}e^{-i\h H\,\paren{t-t_\tx{r}}}
		\paren{e^{i\h H\paren{T_\tx{out}'-t_\tx{r}}}e^{-i\h H_\tx{free}\paren{T_\tx{in}'-t_\tx{r}}}\Ket{\Pi_3\Pi_4}}\nn
	&=	e^{i\h H_\tx{free}\,\paren{t-t_\tx{r}}}e^{-i\h H\,\paren{t-T_\tx{out}'}}
		e^{-i\h H_\tx{free}\paren{T_\tx{out}'-t_\tx{r}}}\Ket{\Pi_3\Pi_4}\nn
	&=	\T e^{-i\int_{T_\tx{out}'}^t\h H^\tx{I}_\tx{int}\fn{t'-t_\tx{r}}\,\df t'}\Ket{\Pi_3\Pi_4}.
}
Just as in the free construction~\eqref{S-matrix in free construction}, we may write the S-matrix as an inner product of the interaction-picture state at an arbitrary time $t$:
\al{
\mc S'
	&=	{}'_\tx{I}\!\Braket{\tx{out; }\Pi_3\Pi_4;t|\tx{in; }\Pi_1\Pi_2;t}_\tx{I}'\nn
	&=	\Bra{\Pi_3\Pi_4}
		\T e^{-i\int_{T_\tx{in}'}^{T_\tx{out}'}\h H_\tx{int}\fn{t'-t_\tx{r}}\,\df t'}
		\Ket{\Pi_1\Pi_2},
		\label{dressed S-matrix}
}
from which the $t$-dependence drops out. Hereafter, we come back to the choice $t_\tx{r}=0$. We note that $\mc S'$ and $\mc S$ are physically different.

If we could take the limits $T_\tx{in}'\to-\infty$ and $T_\tx{out}'\to\infty$, we would be able to write\footnote{
The ``dressed'' construction corresponds to the ordinary plane-wave computation of taking the $T\to \infty\paren{1-i\epsilon}$ limit in $e^{-i\int_{-T}^T\h H^\tx{I}_\tx{int}\fn{t'}\df t'}$ with a positive infinitesimal $\epsilon$, and further switching off the interactions by hand by the replacement $\h H_\tx{int}^\tx{I}\fn{t}\to e^{-\epsilon\ab{t}}\h H_\tx{int}^\tx{I}\fn{t}$ in the S-matrix.
}
\al{
\mc S'
	&\to
		\Bra{\Pi_3\Pi_4}
		\T e^{-i\int_{-\infty}^\infty\h H_\tx{int}\fn{t'}\,\df t'}
		\Ket{\Pi_1\Pi_2}.
}
However, the limits
\al{
T_\tx{in}'
	&\to	-\infty,&
T_\tx{out}'
	&\to	\infty,
	\label{time limits}
}
do not commute with the final-state integral of infinite volume over $\bs\Pi_3$ and $\bs\Pi_4$ as we will see below.

\subsection{Comparison of two constructions}\label{two constructions section}
The in-boundary effect for the fixed configuration $\paren{\bs\Pi_1,\dots,\bs\Pi_4}$  disappears from $\mc S'$, which includes the interaction from the time $T_\tx{in}'$ (or sufficiently earlier time than $T_\tx{in}-\sqrt{2\stout}$ for the given final state configuration) to $T_\tx{in}$ in Eq.~\eqref{replacement of states}.
In the original $\mc S$ in the ``free'' construction, interactions at $t< T_\tx{in}$ does not appear. If we start from $\mc S'$ for the configuration $\paren{\bs\Pi_1,\dots,\bs\Pi_4}$, we recover the boundary effect of $\mc S$ by sharply switching off interactions at $t< T_\tx{in}$.

Here in $\mc S'$, although the free wave packets in $\Ket{\Pi_1\Pi_2}$ are given experimentally at $T_1$ and $T_2$, we identify $\Ket{\Pi_1\Pi_2}$ with the Heisenberg state at much earlier time $T_\tx{in}'$, not at somewhere $T_\tx{in}$ near them. Namely, the Schr\"odinger-picture state $e^{-i\h Ht}\Ket{\tx{in; }\Pi_1\Pi_2}'$ at $t\to T_\tx{in}'$ is identified with the ``free Schr\"odinger-picture'' state $e^{-i\h H_\tx{free}t}\Ket{\Pi_1\Pi_2}$ that is time-evolved backward $t\to T_\tx{in}'$ in a virtual free world governed by $\h H_\tx{free}$, even for the case where interactions are not negligible for $t<T_\tx{in}$. In $\Ket{\tx{in; }\Pi_1\Pi_2}'$, interactions are put at times much earlier than $T_\tx{in}$ at which the supposedly free in-state is to be defined.

For the particular in and out-state configuration $\paren{\bs\Pi_1,\dots,\bs\Pi_4}$ with $\paren{\mf T_\tx{out}-T_\tx{in}}^2\lesssim2\stout$, we may always choose $T_\tx{in}'\ll T_\tx{in}-\sqrt{2\stout}$, and the in-boundary effect for this configuration drops out of $\mc S'$, but there always exist other configuration $(\bs\Pi_3,\bs\Pi_4)$ that has the in-boundary effect at $\mf T_\tx{out}\simeq T_\tx{in}'$ accordingly to Eq.~\eqref{interaction times}.
Therefore, the probability summed over $(\bs\Pi_3,\bs\Pi_4)$ has the in-boundary effect for any fixed $T_\tx{in}'$.

Let us rephrase the above discussion in a slightly different way.
As we move $T_\tx{in}'$ backwards, the bulk region expands, and the effective in-boundary at $T_\tx{in}'$ goes back in time.
For a given $T_\tx{in}'$, the in-boundary contribution arises from the out state that has overlap of out wave packets at $T_\tx{in}'$.
Therefore, the $T_\tx{in}'\to-\infty$ limit is not uniform because the region of in-boundary effect in $\bs\Pi_3\bs\Pi_4$ moves along with $T_\tx{in}'$.
For these out states for given $T_\tx{in}'$, the boundary effect persists. If such an out state is not included, the boundary effect disappears.

To summarize so far, for any configuration of $\bs\Pi_3$ and $\bs\Pi_4$, there always exists a $T_\tx{in}'$ that removes the boundary effect, while for any $T_\tx{in}'$, there always exists a configuration of $\bs\Pi_3$ and $\bs\Pi_4$ that yields an in-boundary effect. Therefore it is subject to debate whether or not the limit~\eqref{time limits} can be taken to remove all the time boundary effects.

The expression for boundary effect in the second term in Eq.~\eqref{illustrative limit of G} vanishes exponentially in the limit $T_\tx{in}\to-\infty$. In the ``dressed'' construction, this is natural because this limit corresponds to taking into account all the interactions from $-\infty$, for the fixed initial and final state configurations. In the ``free'' construction, one emphasizes the fact that no matter how much we take the limit $T_\tx{in}\to-\infty$, there always exists a final state configuration with $\paren{\Re\mc T_\tx{out}-T_\tx{in}}^2\lesssim2\stout$ for a given $T_\tx{in}$. 
The difference of two constructions is the order of procedures: taking the limit $T_\tx{in}\to\infty$ first vs integrating over the infinite volume of $\paren{\bs\Pi_3,\bs\Pi_4}$ first.

So far, both constructions have pros and cons, subject to one's theoretical prejudice.
Ultimately, experiment should determine which (or else) is right.
Currently, an experiment is on-going~\cite{Ushioda:2019hje} based on the ``free'' construction~\cite{Ishikawa:2019nes}.
In this paper, we will leave the choice of constructions open, and concentrate on the wave effect that persists even when we only take into account the bulk effects.
See Sec.~\ref{in-boundary section} for related discussion on the in-boundary effect for $1\to2$ decay of $\Phi\to\phi\phi$.

\section{Bulk amplitude}\label{Bulk amplitude section}
Hereafter, we focus on the bulk contribution and do not take the boundary contributions into account.
We will perform the integration of the virtual momentum $p$ of $\Phi$ in the saddle-point approximation. Note that so far the Gaussian integral over the position of interaction $x$ and $x'$ is exact, up to the time-boundary effects for $t=x^0$ and $t'=x^{\pr0}$.

\subsection{Bulk amplitude after integral over internal momentum}

Neglecting the time-boundary contribution, the probability amplitude in Eq.~\eqref{exact result} becomes
\al{
\mc S	
	&=	i\kappa^2\paren{
			\prod_{A=1}^4{1\ov\sqrt{2E_A}}
			\paren{1\ov\pi\sigma_A}^{3/4}}
		\paren{2\pi\sigma_\tx{in}}^{3/2}
		\paren{2\pi\sigma_\tx{out}}^{3/2}
		\sqrt{2\pi\stin}
		\sqrt{2\pi\stout}
		\int{\df^4p\ov\paren{2\pi}^4}{1\ov p^2+M^2-i\epsilon}\nn
	&\quad\times
		e^{
			-{\sigma_\tx{out}\ov2}\paren{\bs p-\bs P_\tx{out}}^2
			-{\mc R_\tx{out}\ov2}
			+i\ol{\bmf X}_\tx{out}\cdot\paren{\bs p-\bs P_\tx{out}}
			-i\mf T_\tx{out}\paren{p^0-\mc E_\tx{out}-\ol{\bs V}_\tx{out}\cdot\bs p}
			-{\stout\ov2}\paren{p^0-\mc E_\tx{out}-\ol{\bs V}_\tx{out}\cdot\bs p}^2
			}\nn		
	&\quad\times
		e^{
			-{\sigma_\tx{in}\ov2}\paren{\bs p-\bs P_\tx{in}}^2
			-{\mc R_\tx{in}\ov2}
			-i\ol{\bmf X}_\tx{in}\cdot\paren{\bs p-\bs P_\tx{in}}
			+i\mf T_\tx{in}\paren{p^0-\mc E_\tx{in}-\ol{\bs V}_\tx{in}\cdot\bs p}
			-{\stin\ov2}\paren{p^0-\mc E_\tx{in}-\ol{\bs V}_\tx{in}\cdot\bs p}^2
			}.
			\label{master formula}
}
We can square-complete the $p^0$-dependent four terms in the above exponent as
\al{
&-{\varsigma_+\ov2}\paren{p^0-\Omega\fn{\bs p}+i{\delta\mf T\ov\varsigma_+}}^2
		-{\varsigma\ov2}\Paren{\omega_\tx{out}\fn{\bs p}-\omega_\tx{in}\fn{\bs p}}^2
		-{\paren{\delta\mf T}^2\ov2\varsigma_+}\nn
&+i\varsigma\paren{{\mf T_\tx{in}\ov\stin}+{\mf T_\tx{out}\ov\stout}}\paren{\omega_\tx{out}\fn{\bs p}-\omega_\tx{in}\fn{\bs p}}
}
where we have defined
\al{
\varsigma_+
	&:=	\stin+\stout,\nn
\varsigma
	&:=	\paren{{1\ov\stin}+{1\ov\stout}}^{-1},\nn
\delta\mf T
	&:=	\mf T_\tx{out}-\mf T_\tx{in},\nn
\omega_\tx{in}\fn{\bs p}
	&:=	\mc E_\tx{in}+\ol{\bs V}_\tx{in}\cdot\bs p,\nn
\omega_\tx{out}\fn{\bs p}
	&:=	\mc E_\tx{out}+\ol{\bs V}_\tx{out}\cdot\bs p,
}
and the typical ``average energy'' for the $2\to2$ process
\al{
\Omega\fn{\bs p}
	&:=	{\stin\omega_\tx{in}\fn{\bs p}+\stout\omega_\tx{out}\fn{\bs p}\ov\stin+\stout}.
}

By the saddle-point approximation, we get
\al{
\mc S
	&=	i\kappa^2\paren{
			\prod_{A=1}^4{1\ov\sqrt{2E_A}}
			\paren{1\ov\pi\sigma_A}^{3/4}}
		\paren{2\pi\sigma_\tx{in}}^{3/2}
		\paren{2\pi\sigma_\tx{out}}^{3/2}
		\sqrt{2\pi\varsigma}\nn
	&\quad\times
		\int{\df^3\bs p\ov\paren{2\pi}^3}{1\ov-\paren{\Omega\fn{\bs p}-{i\,\delta\mf T\ov\varsigma_+}}^2+\bs p^2+M^2-i\epsilon}\nn
	&\quad\times
		e^{
			-{\sigma_\tx{out}\ov2}\paren{\bs p-\bs P_\tx{out}}^2
			-{\mc R_\tx{out}\ov2}
			+i\ol{\bmf X}_\tx{out}\cdot\paren{\bs p-\bs P_\tx{out}}
			-{\sigma_\tx{in}\ov2}\paren{\bs p-\bs P_\tx{in}}^2
			-{\mc R_\tx{in}\ov2}
			-i\ol{\bmf X}_\tx{in}\cdot\paren{\bs p-\bs P_\tx{in}}
			}\nn
	&\quad\times
		e^{
		-{\varsigma\ov2}\Paren{\omega_\tx{out}\fn{\bs p}-\omega_\tx{in}\fn{\bs p}}^2
		-i\Omega\fn{\bs p}\delta\mf T
		-{\paren{\delta\mf T}^2\ov2\varsigma_+}
		+i\paren{\omega_\tx{out}\fn{\bs p}\mf T_\tx{out}-\omega_\tx{in}\fn{\bs p}\mf T_\tx{in}}
			}.
}
Here, the $\bs p$ dependence of the exponent $e^{\wt F}$ is of the form
\al{
\wt F
	&=	-{\sigma_+\ov2}\bs p^2-{\varsigma\ov2}\paren{\delta\ol{\bs V}\cdot\bs p}^2
+\bs w\cdot\bs p+C,
}
where
\al{
\sigma_+
	&:=	\sigma_\tx{in}+\sigma_\tx{out},\\
\delta\ol{\bs V}
	&:=	\ol{\bs V}_\tx{out}-\ol{\bs V}_\tx{in},\\
\bs w
	&:=	\sigma_+\bs{\mc P}
		-\varsigma\,\delta\mc E\,\delta\ol{\bs V}
		+i\paren{
			\delta\ol{\bmf X}
			+\mf T_\varsigma\,\delta\ol{\bs V}
			},\\
C	&:=	-{\sigma_\tx{in}\ov2}\bs P_\tx{in}^2
		-{\sigma_\tx{out}\ov2}\bs P_\tx{out}^2
		-{\mc R_\tx{in}+\mc R_\tx{out}\ov2}
		-{\varsigma\ov2}\paren{\delta\mc E}^2
		-{\paren{\delta\mf T}^2\ov2\varsigma_+}
			\label{c expression}\nn
	&\quad
		+i\sqbr{
			\ol{\bmf X}_\tx{in}\cdot\bs P_\tx{in}
			-\ol{\bmf X}_\tx{out}\cdot\bs P_\tx{out}
			+\mf T_\varsigma\delta\mc E
			},
}
in which\footnote{
Here we let $\delta$ denote the difference between the in and out quantities in $2\to2$ scattering, rather than the difference between the in and out ones in $1\to2$ decay in Ref.~\cite{Ishikawa:2018koj}.
}
\al{
\delta\mf T
	&:=	\mf T_\tx{out}-\mf T_\tx{in},\\
\delta\ol{\bmf X}
	&:=	\ol{\bmf X}_\tx{out}-\ol{\bmf X}_\tx{in}, \\
\delta\mc E
	&:=	\mc E_\tx{out}-\mc E_\tx{in},
}
and we have defined the ``average momentum'' for the $2\to2$ process
\al{
\bs{\mc P}
	&:=	{\sigma_\tx{in}\bs P_\tx{in}
		+\sigma_\tx{out}\bs P_\tx{out}
			\ov
			\sigma_\tx{in}+\sigma_\tx{out}
			}
}
and the ``interaction time'' for the $2\to2$ process
\al{
\mf T_\varsigma
	&:=	\varsigma\paren{{\mf T_\tx{in}\ov\stin}
			+{\mf T_\tx{out}\ov\stout}}.
}
Note that the last term in Eq.~\eqref{c expression} (in its second line) can be dropped out since it is a pure imaginary constant.

The saddle point ${\p\wt F\ov\p p_i}=0$ is at\footnote{
We have examined the saddle point only looking at the exponential factor.
Around the pole of the propagator, one might need to include its logarithm in the exponent.
}
\al{
p_{*i}
	&=	{w_i\ov \sigma_+}-{\varsigma\paren{\delta\ol{\bs V}}_i\paren{\delta\ol{\bs V}\cdot\bs w}\ov \sigma_+\paren{\sigma_++\varsigma\paren{\delta\ol{\bs V}}^2}},
}
that is,
\al{
\bs p_*
	&=	\paren{
			\bs{\mc P}
			-{\varsigma\,\delta\mc E\,\delta\ol{\bs V}-i\paren{\delta\ol{\bmf X}+\mf T_\varsigma\,\delta\ol{\bs V}}\ov\sigma_+}
			}
		-{\varsigma\paren{\delta\ol{\bs V}}^2\ov\sigma_++\varsigma\paren{\delta\ol{\bs V}}^2}\paren{
			\bs{\mc P}
			-{\varsigma\,\delta\mc E\,\delta\ol{\bs V}-i\paren{\delta\ol{\bmf X}+\mf T_\varsigma\,\delta\ol{\bs V}}\ov\sigma_+}
			}_\parallel
}
where
\al{
\bs Q_\parallel
	&=	{\paren{\delta\ol{\bs V}\cdot\bs Q}\ov\paren{\delta\ol{\bs V}}^2}\delta\ol{\bs V}.
}
Now we can rewrite $\wt F$ without any approximation as
\al{
\wt F
	&=	-{1\ov2}\paren{\bs p-\bs p_*}_i\paren{\sigma_+\delta_{ij}+\varsigma\paren{\delta\ol{\bs V}}_i\paren{\delta\ol{\bs V}}_j}\paren{\bs p-\bs p_*}_j
		+\wt F_*,
}
where
\al{
\wt F_*
	&=	{1\ov2\sigma_+}\paren{
		\bs w^2
		-{\varsigma\paren{\delta\ol{\bs V}\cdot\bs w}^2\ov \sigma_++\varsigma\paren{\delta\ol{\bs V}}^2}
		}
		+C.
}
Let us separate two terms corresponding to the momentum and energy conservation from $\wt F_*$:
\al{
\wt F_*
	&=	F_*
		-{\sigma\ov2}\paren{\delta\bs P}^2
		-{\varsigma\sigma_+\ov2\paren{\sigma_++\varsigma\paren{\delta\ol{\bs V}}^2}}
		\paren{
			\delta E
			-\bs V_\sigma
				\cdot\delta\bs P
			}^2,
}
where we have defined
\al{
\sigma
	&:=	\paren{{1\ov\sigma_\tx{in}}+{1\ov\sigma_\tx{out}}}^{-1}
	=	\paren{\sum_{a=1}^4{1\ov\sigma_a}}^{-1},\\
\delta E
	&:=	E_\tx{out}-E_\tx{in},\\
\delta\bs P
	&:=	\bs P_\tx{out}-\bs P_\tx{in},\\
F_*
	&:=	-{\mc R_\tx{in}+\mc R_\tx{out}\ov2}
		-{\paren{\delta\mf T}^2\ov2\varsigma_+}\nn
	&\phantom{:= \,\,}
		-{\varsigma\ov2\paren{\sigma_++\varsigma\paren{\delta\ol{\bs V}}^2}}
		\paren{
		{\paren{\delta\ol{\bs V}}^2\paren{\delta\ol{\bmf X}}^2
				-\paren{\delta\ol{\bs V}\cdot\delta\ol{\bmf X}}^2
				\ov
				\sigma_+}
		+{\paren{\delta\ol{\bmf X}+\mf T_\varsigma\delta\ol{\bs V}
			}^2\ov\varsigma}
		},
		\label{F star eq}
}
and the ``average velocity'' for the $2\to2$ process
\al{
\bs V_\sigma
	&:=	\sigma\paren{{\ol{\bs V}_\tx{in}\ov\sigma_\tx{in}}+{\ol{\bs V}_\tx{out}\ov\sigma_\tx{out}}},
}
and have used the identity
\al{
\delta\mc E+\delta\ol{\bs V}\cdot\bs{\mc P}=\delta E-\bs V_\sigma\cdot\delta\bs P.
	\label{E identity}
}

We see from the first term in the parentheses in Eq.~\eqref{F star eq}
that the suppression is weaker when the ``impact parameter'' $\delta\ol{\bmf X}$ is parallel to the ``momentum transfer'' $\delta\ol{\bs V}$.
This combination $\paren{\delta\ol{\bs V}}^2\paren{\delta\ol{\bmf X}}^2-\paren{\delta\ol{\bs V}\cdot\delta\ol{\bmf X}}^2$ is always non-negative due to the Cauchy-Schwarz inequality.
Also from the second term, the suppression is weaker when the difference of the average position of in and out states is close at the ``$2\to2$ interaction time'' $\mf T_\varsigma$, namely when $\ab{\delta\ol{\bmf X}+\mf T_\varsigma\delta\ol{\bs V}}$ is small.

For the integrating over $\bs p$, the Gaussian factor is
\al{
\sqrt{\paren{2\pi}^3\ov \sigma_+^2\paren{\sigma_++\varsigma\paren{\delta\ol{\bs V}}^2}}.
}
Finally we get the differential amplitude for a fixed configuration of initial and final states $\paren{\bs\Pi_1,\dots,\bs\Pi_4}$:
\al{
\mc S
	&=	i\mc M
		\paren{
			\prod_{A=1}^4{1\ov\sqrt{2E_A}}
			\paren{1\ov\pi\sigma_A}^{3/4}}
		\nn
	&\quad\times
		\paren{2\pi}^4
		\sqbr{\paren{\sigma\ov2\pi}^{3/2}e^{-{\sigma\ov2}\paren{\delta\bs P}^2}}
		\sqbr{\paren{{1\ov2\pi}{\varsigma\sigma_+\ov\sigma_++\varsigma\paren{\delta\ol{\bs V}}^2}}^{1/2}e^{-{1\ov2}{\varsigma\sigma_+\ov\sigma_++\varsigma\paren{\delta\ol{\bs V}}^2}\paren{\delta E-\bs V_\sigma\cdot\delta\bs P}^2}},
			\label{lengthy result}
}
where we have defined the dimensionless amplitude $\mc M$; cf.\ Eq.~\eqref{S and M}:
\al{
\mc M
	&:=	{\kappa^2e^{F_*}\ov
			-\paren{\paren{\Omega\fn{\bs p_*}}^2-\paren{\delta\mf T\ov\varsigma_+}^2}
			+i2\Omega\fn{\bs p_*}{\delta\mf T\ov\varsigma_+}
			+\bs p_*^2
			+M^2
			-i\epsilon}\nn
	&=	{\kappa^2\ov
			-\paren{\paren{\Omega\fn{\bs p_*}}^2-\paren{\delta\mf T\ov\varsigma_+}^2}
			+i2\Omega\fn{\bs p_*}{\delta\mf T\ov\varsigma_+}
			+\bs p_*^2
			+M^2
			-i\epsilon}\nn
	&\quad\times
		e^{-{\mc R_\tx{in}+\mc R_\tx{out}\ov2}
		-{\paren{\delta\mf T}^2\ov2\varsigma_+}
		-{\varsigma\ov2\paren{\sigma_++\varsigma\paren{\delta\ol{\bs V}}^2}}
		\paren{
		{\paren{\delta\ol{\bs V}}^2\paren{\delta\ol{\bmf X}}^2
				-\paren{\delta\ol{\bs V}\cdot\delta\ol{\bmf X}}^2
				\ov
				\sigma_+}
		+{\paren{\delta\ol{\bmf X}+\mf T_\varsigma\delta\ol{\bs V}
			}^2\ov\varsigma}
		}}.
		\label{amplitude result}
}
Several comments are in order:
\begin{itemize}
\item All the terms in $F_*$ are negative or zero, and hence $F_*$ gives always a suppression factor.
\item In the amplitude~\eqref{lengthy result}, the plane-wave limit $\sigma\to\infty$ gives a delta function for the momentum conservation:
\al{
\paren{\sigma\ov2\pi}^{3/2}e^{-{\sigma\ov2}\paren{\delta\bs P}^2}
	&\to	\delta^3\fn{\delta\bs P}
			=	\delta^3\Fn{\bs P_\tx{out}-\bs P_\tx{in}}.
				\label{momentum delta function}
}
\item Likewise, the limit ${\varsigma\sigma_+\ov\sigma_++\varsigma\paren{\delta\ol{\bs V}}^2}\to\infty$ gives a delta function for the energy conservation:
\al{
\paren{{1\ov2\pi}{\varsigma\sigma_+\ov\sigma_++\varsigma\paren{\delta\ol{\bs V}}^2}}^{1/2}e^{-{1\ov2}{\varsigma\sigma_+\ov\sigma_++\varsigma\paren{\delta\ol{\bs V}}^2}\paren{\delta E-\bs V_\sigma\cdot\delta\bs P}^2}
	&\to	\delta\fn{\delta E-\bs V_\sigma\cdot\delta\bs P}.
	\label{energy delta function}
}
\item In the squared amplitude $\ab{\mc S}^2$, the factor $e^{-\sigma\paren{\delta\bs P}^2}$ gives the momentum conservation in the limit $\sigma\to\infty$:
\al{
\paren{\sigma\ov\pi}^{3/2}e^{-\sigma\paren{\delta\bs P}^2}
	&\to	\delta^3\Paren{\bs P_\tx{out}-\bs P_\tx{in}}.
}
We note that the infinity $\delta^3\fn{0}$ from $\sqbr{\delta^3\Paren{\delta\bs P}}^2$ that appears in the plane-wave computation, using the right-hand side in Eq.~\eqref{momentum delta function}, is tamed in the current wave-packet one: The would-be delta function squared becomes another would-be delta function again.
\item Likewise, the factor
\als{
\exp\fn{
	-{\varsigma\sigma_+\ov\sigma_++\varsigma\paren{\delta\ol{\bs V}}^2}
	\paren{\delta E-\bs V_\sigma\cdot\delta\bs P}^2}
}
in $\ab{\mc S}^2$ gives the energy conservation in the limit ${\varsigma\sigma_+\ov\sigma_++\varsigma\paren{\delta\ol{\bs V}}^2}\to\infty$:
\al{
\sqrt{{1\ov\pi}{\varsigma\sigma_+\ov\sigma_++\varsigma\paren{\delta\ol{\bs V}}^2}}\,
e^{
	-{\varsigma\sigma_+\ov\sigma_++\varsigma\paren{\delta\ol{\bs V}}^2}
	\paren{\delta E-\bs V_\sigma\cdot\delta\bs P}^2
	}
&\to	\delta\Paren{E_\tx{out}-E_\tx{in}-\bs V_\sigma\cdot\paren{\bs P_\tx{out}-\bs P_\tx{in}}}.
}
Note that the energy conservation is deformed by the wave-packet effect $\bs V_\sigma\cdot\delta\bs P$, which goes to zero in the momentum conserving limit: $\delta\bs P\to0$.
\item It is remarkable that the wave effect persists even without the time-boundary effect. Namely, the real and imaginary parts of the pole of propagator are shifted as in Eq.~\eqref{amplitude result}. Even when $\bs p_*\simeq P_\tx{in}$ and $\Omega\fn{\bs p_*}\simeq E_\tx{in}$, the pole position of the propagator is shifted such that the mass-squared $M^2$ and decay width $\Gamma$ are shifted by $\paren{\delta\mf T/\varsigma_+}^2$ and $-2E_\tx{in}\delta\mf T/\varsigma_+M$, respectively.
\end{itemize}

\subsection{In-boundary effect for decay}
\label{in-boundary section}

Here we discuss how our result for the $2\to2$ scattering $\phi\phi\to\Phi\to\phi\phi$ can be applied to the $1\to2$ decay process $\Phi\to\phi\phi$.
In Sec.~\ref{boundary effect section}, we have presented two different constructions regarding the boundary effect.
For the $1\to2$ decay $\Phi\to\phi\phi$~\cite{Ishikawa:2018koj}, the key question for its in-boundary effect is how we can better take into account the production process of $\Phi$. Which approximates an experimentally prepared state of $\Phi$ better at an initial time $T_\tx{in}^\tx{decay}$? Is it the Heisenberg state
\al{
\Ket{\tx{in; }\Phi}=e^{i\h HT_\tx{in}^\tx{decay}}e^{-i\h H_\tx{free}T_\tx{in}^\tx{decay}}\Ket{\Phi}
}
in the free construction, or
\al{
\Ket{\tx{in; }\Phi}'
	&=	e^{i\h HT_\tx{in}^\tx{decay}}e^{-i\h H_\tx{free}T_\tx{in}^\tx{decay}}\T e^{-i\int_{T'}^{T_\tx{in}^\tx{decay}}H^\tx{I}_\tx{int}\fn{t'}\,\df t'}\Ket{\Phi}&
	&(T'\to-\infty)
}
in the dressed construction?\footnote{
See the discussion in Secs.~\ref{dressed section} and \ref{two constructions section} for subtleties on taking $T'\to-\infty$ limit.
}

In our result for the $2\to2$ $s$-channel scattering of $\phi\phi\to\Phi\to\phi\phi$, the interaction time $\mf T_\tx{in}$ would correspond to $T_\tx{in}^\tx{decay}$ for the $\Phi\to\phi\phi$ decay.
Here we note that the in-boundary effect of the decay becomes significant when the decay-interaction point around $\mf T_\tx{out}$ is near the center of the in-state wave packet at $T_\tx{in}^\tx{decay}\simeq\mf T_\tx{in}$, namely when
\al{
\paren{\delta\mf T}^2=\paren{\mf T_\tx{out}-\mf T_\tx{in}}^2\lesssim 2\stout.
}
Therefore, one might interpret that the limit $\delta\mf T\to0$, which necessarily arises when we integrate over the final state phase space $\bs\Pi_3$ and $\bs\Pi_4$, corresponds to the in-boundary for the $1\to2$ decay.
By taking $\delta\mf T\to0$ in Eq.~\eqref{amplitude result}, we obtain
\al{
\mc M
	&\to
		{\kappa^2\ov
			-\Paren{\Omega\fn{\bs p_*}}^2
			+\bs p_*^2
			+M^2
			-i\epsilon}\nn
	&\quad\times
		e^{-{\mc R_\tx{in}+\mc R_\tx{out}\ov2}
		-{\varsigma\ov2\paren{\sigma_++\varsigma\paren{\delta\ol{\bs V}}^2}}
		\paren{
		{\paren{\delta\ol{\bs V}}^2\paren{\delta\ol{\bmf X}}^2
				-\paren{\delta\ol{\bs V}\cdot\delta\ol{\bmf X}}^2
				\ov
				\sigma_+}
		+{\paren{\delta\ol{\bmf X}+\mf T_\varsigma\delta\ol{\bs V}
			}^2\ov\varsigma}
		}}.
}
We see that there is no $1\to2$ in-boundary effect in the $2\to2$ bulk amplitude.
If the in-boundary effect of $1\to2$ decay exists, it can only emerge from the in-boundary effect of $2\to2$ scattering.


\section{Various limits}\label{limits section}

Here, we take several limits where $\sigma_\tx{in}$ and/or $\sigma_\tx{out}$ goes to infinity.

\subsection{Plane-wave limit for initial state}
First we take the plane-wave limit for the initial state $\sigma_\tx{in} \to \infty$ for fixed $\sigma_\tx{out}$:
\al{
\sigma
	= {{\sigma_\tx{out}}\ov{1 + {\sigma_\tx{out}\ov\sigma_\tx{in}}}}  
	&\to  \sigma_\tx{out},\\
\varsigma
	= {{\stout}\ov{1 + {\stout\ov\stin}}}  
	&\to  \stout,\\
{{\varsigma\sigma_+}\ov{\sigma_+ + \varsigma \paren{\delta\ol{\bs V}}^2}}
	= {{\sigma}\ov{  {\sigma_\tx{in}\ov\sigma_+} \Delta\bs V_\tx{out}^2 + {\sigma_\tx{out}\ov\sigma_+} \Delta\bs V_\tx{in}^2 
		+ {\sigma\ov\sigma_+} \paren{\delta\ol{\bs V}}^2 }}
	&\to \stout,\\
\varsigma\ov\sigma_+
	&\to 0,
}
where, since $\sigma$ and ${{\varsigma\sigma_+}\ov{\sigma_+ + \varsigma \paren{\delta\ol{\bs V}}^2}}$ stay finite,
both of the momentum and energy conservations are violated.
The above limited values lead to
\al{
\bs{\mc P}
	=	{\bs P_\tx{in}
		+{\sigma_\tx{out}\ov\sigma_\tx{in}}\bs P_\tx{out}
			\ov
			{1+{\sigma_\tx{out}\ov\sigma_\tx{in}}}
			}
	&\to \bs P_\tx{in},
		\label{mcPform-in-limit-I} \\
\bs p_\ast
	&\to 	\bs{\mc P}
			=
			\bs{P}_\text{in},
		\label{pastform-in-limit-I} \\
\Omega\fn{\bs p_\ast}
	=	{\omega_\tx{in}\fn{\bs p_\ast}+{\stout\ov\stin}\omega_\tx{out}\fn{\bs p_\ast}\ov{1+{\stout\ov\stin}}}
	&\to \omega_\tx{in}\fn{\bs p_\ast} 
	= 	\omega_\tx{in}\fn{\bs P_\tx{in}}
	=	E_\tx{in}, 
		\label{Omegaform-in-limit-I} \\
\mc R_\tx{in}
	&\to	0,\\
F_*
	&\to 	-{\mc R_\tx{out}\ov2},\\
\bs V_\sigma
	&\to	\ol{\bs V}_\tx{out},
}
where we used the result of Eq.~\eqref{mcPform-in-limit-I} in the last steps of Eqs.~\eqref{pastform-in-limit-I} 
and \eqref{Omegaform-in-limit-I}.
From the above information, we get the limit of propagator
\al{
{1\ov
			-\paren{\paren{\Omega\fn{\bs p_*}}^2-\paren{\delta\mf T\ov\varsigma_+}^2}
			+i2\Omega\fn{\bs p_*}{\delta\mf T\ov\varsigma_+}
			+\bs p_*^2
			+M^2
			-i\epsilon} 
\to  {1\ov
			-E_\tx{in}^2
			+\bs{P}_\text{in}^2
			+M^2
			-i\epsilon}.
}
To summarize,
\al{
\mc S
	&\to
		i
		\paren{
			\prod_{A=1}^4{1\ov\sqrt{2E_A}}
			\paren{1\ov\pi\sigma_A}^{3/4}}
		{\kappa^2\ov-E_\tx{in}^2+\bs{P}_\text{in}^2+M^2-i\epsilon}e^{-{\mc R_\tx{out}\ov2}}
		\nn
	&\quad\times
		\paren{2\pi}^4
		\sqbr{\paren{\sigma_\tx{out}\ov2\pi}^{3/2}e^{-{\sigma_\tx{out}\ov2}\paren{\delta\bs P}^2}}
		\sqbr{\paren{\stout\ov2\pi}^{1/2}e^{-{\stout\ov2}\paren{\delta E-\ol{\bs V}_\tx{out}\cdot\delta\bs P}^2}}.
}
We see that the momentum conservation is broken by $\sim\sqrt{\sigma_\tx{out}}$, and the energy conservation by $\sim\sqrt{\stout}$, along with the shift $-\ol{\bs V}_\tx{out}\cdot\delta\bs P$ in the plane-wave limit for the initial state.

\subsection{Plane-wave limit for final state}
Similarly, we may take the plane-wave limit for the final state $\sigma_\tx{out} \to \infty$ for fixed $\sigma_\tx{in}$:
\al{
\sigma
	&\to  \sigma_\tx{in},\\
\varsigma
	&\to  \stin,\\
{{\varsigma\sigma_+}\ov{\sigma_+ + \varsigma \paren{\delta\ol{\bs V}}^2}}
	&\to \stin,\\
\varsigma\ov\sigma_+
	&\to 0,  \\
\bs{\mc P}
	&\to 	 \bs P_\tx{out}, \\
\bs p_\ast
	&\to 	\bs{P}_\text{out},\\
\Omega\fn{\bs p_\ast}
	&\to \omega_\tx{out}\fn{\bs P_\tx{out}}
	=	E_\tx{out}, \\
\mc R_\tx{out}
	&\to	0,\\
F_*
	&\to 	-{\mc R_\tx{in}\ov2},\\
\bs V_\sigma
	&\to	\ol{\bs V}_\tx{in}.
}
The limit of propagator becomes
\al{
{1\ov
			-\paren{\paren{\Omega\fn{\bs p_*}}^2-\paren{\delta\mf T\ov\varsigma_+}^2}
			+i2\Omega\fn{\bs p_*}{\delta\mf T\ov\varsigma_+}
			+\bs p_*^2
			+M^2
			-i\epsilon} 
\to  {1\ov
			-E_\tx{out}^2
			+\bs{P}_\text{out}^2
			+M^2
			-i\epsilon}.
}
To summarize,
\al{
\mc S
	&\to
		i
		\paren{
			\prod_{A=1}^4{1\ov\sqrt{2E_A}}
			\paren{1\ov\pi\sigma_A}^{3/4}}
		{\kappa^2\ov-E_\tx{out}^2+\bs{P}_\text{out}^2+M^2-i\epsilon}e^{-{\mc R_\tx{in}\ov2}}
		\nn
	&\quad\times
		\paren{2\pi}^4
		\sqbr{\paren{\sigma_\tx{in}\ov2\pi}^{3/2}e^{-{\sigma_\tx{in}\ov2}\paren{\delta\bs P}^2}}
		\sqbr{\paren{\stin\ov2\pi}^{1/2}e^{-{\stin\ov2}\paren{\delta E-\ol{\bs V}_\tx{in}\cdot\delta\bs P}^2}}.
}
We see that the momentum conservation is broken by $\sim\sqrt{\sigma_\tx{in}}$, and the energy conservation by $\sim\sqrt{\stin}$, along with the shift $-\ol{\bs V}_\tx{in}\cdot\delta\bs P$ in the plane-wave limit for final state.

\subsection{Plane-wave limit for both}
Finally, we take the double-scaling limit $\sigma_\tx{in},\,\sigma_\tx{out} \to \infty$ for fixed $\sigma_\tx{out}/\sigma_\tx{in}$:
\al{
\sigma
	= {{\sigma_\tx{out}}\ov{1 + {\sigma_\tx{out}\ov\sigma_\tx{in}}}}  
	&\to  \infty,	\label{momentum conservation limit}\\
\varsigma
	= {{\stout}\ov{1 + {\stout\ov\stin}}}  
	&\to  \infty,\\
{{\varsigma\sigma_+}\ov{\sigma_+ + \varsigma \paren{\delta\ol{\bs V}}^2}}
	= {{\sigma}\ov{  {\sigma_\tx{in}\ov\sigma_+} \Delta\bs V_\tx{out}^2 + {\sigma_\tx{out}\ov\sigma_+} \Delta\bs V_\tx{in}^2 
		+ {\sigma\ov\sigma_+} \paren{\delta\ol{\bs V}}^2 }}
	&\to \infty,
		\label{energy conservation limit}\\
\varsigma\ov\sigma_+
	&\to
		{{\sigma_\tx{out}/\sigma_\tx{in}}\ov
		{\paren{1 + {\sigma_\tx{out}\ov\sigma_\tx{in}}}  
			\paren{\Delta\bs V_\tx{out}^2 +  \Delta\bs V_\tx{in}^2{\sigma_\tx{out}\ov\sigma_\tx{in}}} }}.
}

The limits~\eqref{momentum conservation limit} and \eqref{energy conservation limit} lead to the momentum and energy conserving delta functions $\delta^3\fn{\bs{P}_\text{out}-\bs{P}_\text{in}}$ and $\delta\fn{E_\tx{out}-E_\tx{in}}$ as in Eqs.~\eqref{momentum delta function} and \eqref{energy delta function}, respectively.
Then we obtain
\al{
\delta\mc E
	&\approx
		-\delta\ol{\bs V}\cdot\bs{\mc P},\\
\bs{\mc P}
	&=	{\bs P_\tx{in}
		+{\sigma_\tx{out}\ov\sigma_\tx{in}}\bs P_\tx{out}
			\ov
			{1+{\sigma_\tx{out}\ov\sigma_\tx{in}}}
			}
	\approx \bs P_\tx{in} \approx \bs P_\tx{out}, \\
\bs p_\ast
	&\to
	 	\paren{
			\bs{\mc P}
			-{\varsigma\,\delta\mc E\,\delta\ol{\bs V}\ov\sigma_+}
			}
		-{{\varsigma\ov\sigma_+}\paren{\delta\ol{\bs V}}^2\ov1+{\varsigma\ov\sigma_+}\paren{\delta\ol{\bs V}}^2}
		\paren{
			\bs{\mc P}
			-{\varsigma\,\delta\mc E\,\delta\ol{\bs V}\ov\sigma_+}
			}_\parallel\nn
	&\approx
		\bs{\mc P},\\
\Omega\fn{\bs p_\ast}
	&=	{\sqbr{E_\tx{in}-\ol{\bs V}_\tx{in}\cdot \paren{\bs P_\tx{in} - \bs p_\ast}}
		+{\stout\ov\stin} \sqbr{E_\tx{out}-\ol{\bs V}_\tx{out}\cdot \paren{\bs P_\tx{out} - \bs p_\ast}}
			\ov{1+{\stout\ov\stin}}}\nn
	&\approx
		{E_\tx{in}
		+{\stout\ov\stin} E_\tx{out}
			\ov{1+{\stout\ov\stin}}}
	\approx
		E\fn{\bs{\mc P}}
	\approx
		E_\tx{in}
	\approx
		E_\tx{out}, \\
F_*
	&\to 	0,
}
where $\approx$ denotes that we have used the energy and momentum conservation from the above mentioned delta functions.
Based on the above information, we derive the plane-wave limit of the propagator:
\al{
{1\ov
			-\paren{\paren{\Omega\fn{\bs p_*}}^2-\paren{\delta\mf T\ov\varsigma_+}^2}
			+i2\Omega\fn{\bs p_*}{\delta\mf T\ov\varsigma_+}
			+\bs p_*^2
			+M^2
			-i\epsilon} 
&\to
	 {1\ov
			-\paren{\Omega\fn{\bs{p}_\ast}}^2
			+\bs p_*^2
			+M^2
			-i\epsilon}\nn
&\approx
		 {1\ov
			-\Paren{E\fn{\bs{\mc P}}}^2
			+\bs{\mc P}^2
			+M^2
			-i\epsilon}.
}
We see that the propagator is reduced to the plane-wave form.
To summarize,
\al{
\mc S
	&\to
		i
		\paren{
			\prod_{A=1}^4{1\ov\sqrt{2E_A}}
			\paren{1\ov\pi\sigma_A}^{3/4}}
		{\kappa^2\ov
			-\Paren{E\fn{\bs{\mc P}}}^2
			+\bs{\mc P}^2
			+M^2
			-i\epsilon}
		\paren{2\pi}^4
		\delta^4\fn{P_\tx{out}-P_\tx{in}},
}
where $\delta^4\fn{P_\tx{out}-P_\tx{in}}=\delta\fn{E_\tx{out}-E_\tx{in}}\delta^3\fn{\bs P_\tx{out}-\bs P_\tx{in}}$.

\section{Discussion}\label{summary section}
In this paper, we have computed the Gaussian S-matrix for the $s$-channel $2\to2$ scalar scattering: $\phi\phi\to\Phi\to\phi\phi$.
We have found that the wave effects persist even without the time-boundary effect.

As a future work, it would be interesting to study the integrated probability  after performing the final state integral over the positions $\bs X_3$ and $\bs X_4$:
\al{
\int\df^3\bs X_3\,\df^3\bs X_4\ab{\mc S}^2.
}
Then we may read off how the ordinary plane-wave differential cross section arises, and see the derivation from it due to the wave effects.
It would also be interesting to study the factorization in the limit $\paren{E_\tx{in}^2-\bs P_\tx{in}^2}\to M^2$.

\subsection*{Acknowledgment}
We thank Hiromasa Nakatsuka for useful discussion and Referee for careful reading of the manuscript.
The work of K.O.\ is in part supported by JSPS Kakenhi Grant No.~19H01899.

\appendix
\section*{Appendix}

\section{Comparison with $\phi^4$ theory}\label{phi 4 section}
Let us consider an interaction Hamiotonian
\al{
\h H_\tx{int}\fn{t}
	&=	{\lambda\ov4!}\int\df^3\bs x\,\h\phi^4\fn{x}.
}
The the tree-level probability amplitude becomes
\al{
\mc S
	&=	-i\lambda
		\int_{T_\tx{in}}^{T_\tx{out}}\df t\int\df^3\bs x\,
		f^*_{\sigma_3;\Pi_3}\fn{x}
		f^*_{\sigma_4;\Pi_4}\fn{x}
		f_{\sigma_1;\Pi_1}\fn{x}
		f_{\sigma_2;\Pi_2}\fn{x}.
}
In the leading plane-wave approximation, we get
\al{
\mc S
	&\to
		-i\lambda\paren{\prod_{A=1}^4\paren{1\ov\pi\sigma_A}^{3/4}{1\ov\sqrt{2E_A}}}
		\int_{T_\tx{in}}^{T_\tx{out}}\df t\int\df^3\bs x\nn
	&\quad
		\times e^{iP_1\cdot\paren{x-X_1}-{\paren{\bs x-\bs\Xi_1\fn{t}}^2\ov2\sigma_1}}
		e^{iP_2\cdot\paren{x-X_2}-{\paren{\bs x-\bs\Xi_2\fn{t}}^2\ov2\sigma_2}}
		e^{-iP_3\cdot\paren{x-X_3}-{\paren{\bs x-\bs\Xi_3\fn{t}}^2\ov2\sigma_3}}
		e^{-iP_4\cdot\paren{x-X_4}-{\paren{\bs x-\bs\Xi_4\fn{t}}^2\ov2\sigma_4}}\nn
	&=	-i\lambda\int_{T_\tx{in}}^{T_\tx{out}}\df t
		\paren{\prod_{A=1}^4\paren{1\ov\pi\sigma_A}^{3/4}{1\ov\sqrt{2E_A}}e^{i\alpha_AE_A\paren{t-T_A}}}\nn
	&\quad\times
		\int\df^3\bs x\,
		e^{\sum_{A=1}^4\paren{-{\paren{\bs x-\bs\Xi_A\fn{t}}^2\ov2\sigma_A}-i\alpha_A\bs P_A\cdot\paren{\bs x-\bs X_A}}}\nn
	&=	-i\lambda\int_{T_\tx{in}}^{T_\tx{out}}\df t
		\paren{\prod_{A=1}^4\paren{1\ov\pi\sigma_A}^{3/4}{1\ov\sqrt{2E_A}}e^{i\alpha_AE_A\paren{t-T_A}}}\nn
	&\quad\times
		\int\df^3\bs x\,
		e^{-{1\ov2\sigma}\ol{\paren{\bs x-\bs\Xi\fn{t}}^2}
			-i\,\ol{\alpha\sigma\bs P\cdot\paren{\bs x-\bs X}}},
}
where $\alpha_1=\alpha_2=-1$, $\alpha_3=\alpha_4=1$, and $\sigma:=\paren{\sum_{A=1}^4{1\ov\sigma_A}}^{-1}$.
The exponent becomes
\al{
\tx{exponent}
	&=	-{1\ov2\sigma}\paren{\bs x-\ol{\bs\Xi\fn{t}}+i\sigma\,\delta\bs P}^2
		-{1\ov2\sigma}\paren{
			\ol{\paren{\bs\Xi\fn{t}}^2}
			-\paren{\ol{\bs\Xi\fn{t}}}^2
			}\nn
	&\phantom{= \,\,}
		-{\sigma\ov2}\paren{\delta\bs P}^2
		-i\,\ol{\bs\Xi\fn{t}}\cdot\delta\bs P
		+i\,\delta E\,t
		+i\sqbr{\cdots},
}
where $+i\sqbr{\cdots}$ denotes irrelevant imaginary constant terms which disappear in $\ab{S}^2$ and we have used $\ol{\alpha\sigma E}=\sigma\,\delta E$ and $\ol{\alpha\sigma\bs P}=\sigma\,\delta\bs P$.
Now
\al{
\Delta\paren{\bs\Xi\fn{t}}^2
	&=	\ol{\paren{\bs\Xi\fn{t}}^2}
			-\paren{\ol{\bs\Xi\fn{t}}}^2\nn
	&=	\ol{\paren{\bmf X+\bs Vt}^2}
		-\paren{\ol{\bmf X+\bs Vt}}^2\nn
	&=	\ol{\bmf X^2}-\ol{\bmf X}^2
		+\paren{\ol{\bs V^2}-\ol{\bs V}^2}t^2
		-2\paren{\ol{\bmf X}\cdot\ol{\bs V}-\ol{\bmf X\cdot\bs V}}t\nn
	&=	\Delta\bmf X^2+\Delta\bs V^2\,t^2
		-2\paren{\ol{\bmf X}\cdot\ol{\bs V}-\ol{\bmf X\cdot\bs V}}t.
}
After integrating over $\bs x$, the exponent becomes
\al{
\tx{exponent}
	&=	-{\Delta\bmf X^2\ov2\sigma}
		-{\Delta\bs V^2\ov2\sigma}\paren{
			t-{\ol{\bmf X}\cdot\ol{\bs V}-\ol{\bmf X\cdot\bs V}\ov\Delta\bs V^2}
			}^2
		+{\Delta\bs V^2\ov2\sigma}
		\paren{\ol{\bmf X}\cdot\ol{\bs V}-\ol{\bmf X\cdot\bs V}\ov\Delta\bs V^2}^2
		\nn
	&\quad
		-{\sigma\ov2}\paren{\delta\bs P}^2
		-i\,\ol{\bmf X+\bs Vt}\cdot\delta\bs P
		+i\,\delta E\,t
		+i\sqbr{\cdots}\nn
	&=	-{\Delta\bs V^2\ov2\sigma}\paren{
			t
			-{\ol{\bmf X}\cdot\ol{\bs V}-\ol{\bmf X\cdot\bs V}
			-i\sigma\paren{\delta E-\ol{\bs V}\cdot\delta\bs P}
			\ov\Delta\bs V^2}
			}^2\nn
	&\quad
		-{1\ov2}\paren{
			{\Delta\bmf X^2\ov\sigma}
			-{\Delta\bs V^2\ov\sigma}
				\paren{\ol{\bmf X}\cdot\ol{\bs V}-\ol{\bmf X\cdot\bs V}
					\ov\Delta\bs V^2}^2
			}
		\nn
	&\quad
		-{\sigma\paren{\delta E-\ol{\bs V}\cdot\delta\bs P}^2
				\ov
				2\Delta\bs V^2}
		-{\sigma\ov2}\paren{\delta\bs P}^2
		+i\sqbr{\cdots}.
}
In the last expression, the second term corresponds to the overlap exponent $-\mc R/2$, with $\mc R={\Delta\bmf X^2\ov\sigma}-{\mf T^2\ov\varsigma}$ being non-negative (see Sec.~3.1 in Ref.~\cite{Ishikawa:2018koj}), and the third and fourth terms to the energy and momentum conservations, respectively.

After integrating over $\bs x$ and $t$ (neglecting the time-boundaries), we get the expression for the probability amplitude, namely the dimensionless $\mc S$-matrix:
\al{
\mc S
	&=	-i\lambda\paren{2\pi\sigma}^{3/2}\sqrt{2\pi\varsigma}
		\paren{
			\prod_{A=1}^4\paren{1\ov\pi\sigma_A}^{3/4}{1\ov\sqrt{2E_A}}
			}
		e^{-{\mc R\ov2}-{\sigma\ov2}\paren{\delta\bs P}^2-{\varsigma\ov2}\paren{\delta E-\ol{\bs V}\cdot\delta\bs P}^2}\nn
	&=	-i\lambda\paren{2\pi}^4
		\paren{
			\prod_{A=1}^4\paren{1\ov\pi\sigma_A}^{3/4}{1\ov\sqrt{2E_A}}
			}
		e^{-{\mc R\ov2}}
		\paren{\paren{\sigma\ov2\pi}^{3/2}e^{-{\sigma\ov2}\paren{\delta\bs P}^2}}
		\paren{
			\sqrt{\varsigma\ov2\pi}
			e^{-{\varsigma\ov2}\paren{\delta E-\ol{\bs V}\cdot\delta\bs P}^2}
			}.
}
We may compare this result with the relation between the dimensionful plane-wave S-matrix element $S_\tx{plane}$ and the dimensionless plane-wave amplitude $\mc M_\tx{plane}$:
\al{
S_\tx{plane}=i\paren{2\pi}^4\delta^4\fn{P_\tx{out}-P_\tx{in}}\mc M_\tx{plane}.
}
We see that
\al{
\mc M
	&=	\mc M_\tx{plane} \, e^{-\mc R/2}
}
gives the proper normalization, where $\mc M_\tx{plane}=-\lambda$ for the current case.
That is,
\al{
\mc S
	&=	i\mc M\paren{2\pi}^4
		\paren{
			\prod_{A=1}^4\paren{1\ov\pi\sigma_A}^{3/4}{1\ov\sqrt{2E_A}}
			}
		\paren{\paren{\sigma\ov2\pi}^{3/2}e^{-{\sigma\ov2}\paren{\delta\bs P}^2}}
		\paren{
			\sqrt{\varsigma\ov2\pi}
			e^{-{\varsigma\ov2}\paren{\delta E-\ol{\bs V}\cdot\delta\bs P}^2}
			}.
		\label{S and M}
}

\bibliographystyle{utphys}
\bibliography{refs}

\end{document}